\documentclass[12pt]{article}
\usepackage{epsfig,amssymb}

\newcommand{\bq}{\begin{equation}}
\newcommand{\eq}{\end{equation}}
\newcommand{\bqa}{\begin{eqnarray}}
\newcommand{\eqa}{\end{eqnarray}}
\newcommand{\ben}{\begin{enumerate}}
\newcommand{\een}{\end{enumerate}}
\newcommand{\bc}{\begin{center}}
\newcommand{\ec}{\end{center}}
\newcommand{\bqb}{\begin{eqnarray*}}
\newcommand{\eqb}{\end{eqnarray*}}

\def\gsim{\gtrsim}

\def\pr#1#2#3{ Phys. Rev. ${\bf{#1}}$ (#2) #3}

\def\pl#1#2#3{ Phys. Lett. ${\bf{#1}}$ (#2) #3}
\def\prep#1#2#3{ Phys. Rep. ${\bf{#1}}$ (#2) #3}

\def\np#1#2#3{ Nucl. Phys. ${\bf{#1}}$ (#2) #3}
\def\zp#1#2#3{ Z. f. Phys. ${\bf{#1}}$ (#2) #3}

\def\cpc#1#2#3{ Comput. Phys. Commun. ${\bf{#1}}$ (#2) #3}

\def\ie{{\it i.e.\/}}
\def\eg{{\it e.g.\/}}

\def\etal{{\it et.al.\/}}

\global\nulldelimiterspace = 0pt

\def\L{ {\cal L }}

\def\C{ {\cal C }}
\def\D{ {\cal D }}
\def\R{ {\cal R }}

\def\mwd{m_W^2}
\def\mw{m_W}

\def\slepton{\tilde l}
\def\msl{M_{\tilde l}}
\def\msld{M_{\tilde l}^2}

\def\t{\hat t}
\def\s{\hat s}
\def\u{\hat u}

\begin{document}
\pagenumbering{arabic}
\thispagestyle{empty}
\def\thefootnote{\fnsymbol{footnote}}
\setcounter{footnote}{1}

\begin{flushright}
PM/99-04 \\
THES-TP 99/01 \\
hep-ph/......... \\
January 1999
 \end{flushright}
\vspace{2cm}
\begin{center}
{\Large\bf The $\gamma \gamma \to \gamma \gamma $ process in
the Standard and  SUSY models at high energies.}\footnote{Partially
supported by the NATO grant CRG 971470 and
by the Greek government grant PENED/95 K.A. 1795.}
 \vspace{1.5cm}  \\
{\large G.J. Gounaris$^a$, P.I. Porfyriadis$^a$ and
F.M. Renard$^b$}\\
\vspace{0.7cm}
$^a$Department of Theoretical Physics, University of Thessaloniki,\\
Gr-54006, Thessaloniki, Greece.\\
\vspace{0.2cm}
$^b$Physique
Math\'{e}matique et Th\'{e}orique,
UMR 5825\\
Universit\'{e} Montpellier II,
 F-34095 Montpellier Cedex 5.\\
\vspace{0.2cm}

\vspace*{1cm}

{\bf Abstract}
\end{center}

We study the helicity amplitudes of the process
$\gamma \gamma \to \gamma \gamma$ at high energy, which
in the standard  and SUSY models first arise at the
one-loop order. In the standard model (SM), the diagrams involve
$W$ and charged quark and lepton loops, while in SUSY we also have
contributions from  chargino, charged sfermion and Higgs loop
diagrams. The SUSY contributions are most important in the region
above the threshold for producing the supersymmetric
partners; since there, they interfere most effectively with
the primarily imaginary SM amplitudes. Simple expressions for
the relevant 1-loop functions are given, which provide a
direct overview of the behaviour of the helicity amplitudes
in the whole parameter space at high energies.
The various characteristics of a large set of
observables are studied in detail.

\def\thefootnote{\arabic{footnote}}
\setcounter{footnote}{0}
\clearpage

\section{Introduction}

A striking option  for a the future $e^+e^-$ Linear
Collider (LC) \cite{LC, LC-PhysRep}, is to operate it as a
$\gamma \gamma $
Collider ($LC_{\gamma \gamma}$) whose  c.m. energy
may be variable  and as high as $80\%$ of the initial $e^+e^-$
c.m. energy \cite{LCgg}. According to the present ideas, this should
be achieved by  colliding  each of the
$e^\pm$ beams with laser photons,
which are subsequently backscattered, through the Compton effect.
This way, very energetic photons along the
$e^\pm$ direction are generated, while $e^\pm$ loose most of their
energy  \cite{LCgg, gamma97}. The energy spectrum and spin
composition of the two photon beams, in the thus generated $\gamma \gamma $
Collider, depend of course on the energies and polarization
conditions of the $e^\pm$ beams and lasers.
At present, there are still many technical details to overcome, before
deciding that such an option is really viable \cite{gamma97}.
In this respect, it is necessary to assess its  importance,
before deciding   whether the physics
opportunities  there, justify the effort. \par

Up to now it has been seen in many cases that  $LC_{\gamma \gamma}$
is more
powerful than LC, in searching for New Physics (NP) beyond the
Standard Model (SM); mainly because the $\gamma \gamma $ initial state
has the tendency to couple stronger than the $e^+e^-$ one, to the new
degrees of freedom contained in many forms of NP
\cite{gamma-V, gamma-H}. Such searches may involve either the direct
production of new degrees of freedom (like \eg\@ charginos, light sleptons or
 a light $\tilde t_1$ (stop) in  SUSY models) \cite{Zerwas-SUSY};
or the precise study of the production of  SM particles
like \eg~ in $\gamma \gamma \to W^+W^-,~H$ or the production of
Higgs pairs,
where the new degrees of freedom contribute virtually,
in some loop diagrams  \cite{LC, gamma-V, gamma-H, Zerwas-SUSY}. \par

In this respect, processes like $\gamma \gamma \to \gamma \gamma $,
$\gamma \gamma \to Z\gamma $, $\gamma \gamma \to ZZ$ should also provide
very important tools for searching or constraining  NP;
particularly because the SM contribution there, first appears at
the 1-loop level and should be small. In the present paper,
 we concentrate on the $\gamma \gamma \to \gamma \gamma $ process,
which in SM is fully determined by the contributions of
charged fermion   and $W$ loops. The $W$-loop contribution has
been first calculated in \cite{Jikia} in terms of the standard  1-loop
functions of \cite{Passarino}, while the  expression for
the fermion contribution in terms of the same functions
has been given in \cite{vanderBij}.  \par

The structure of the $\gamma \gamma \to \gamma \gamma $ helicity
amplitudes  at high energies ($\sqrt{s_{\gamma \gamma}} \gsim 0.3~ TeV$)
and {\it any} scattering angle, turns out to be remarkably simple and
intuitive.
In the Standard Model, the whole process is dominated at
high energies, by the
helicity {\it non-flip} amplitudes $F_{\pm\pm\pm\pm}(s,t,u)$
and\footnote{This equality is due to Bose statistics.}
$F_{\pm\mp\pm\mp}(s,t,u)=F_{\pm\mp\mp\pm}(s,u,t)$,
which are predominantly imaginary  for all scattering angles
\cite{Jikia,GPRgamma1}.
The dominant contribution to these amplitudes for
$\sqrt{s_{\gamma \gamma}} \gsim 0.3~TeV$, is easily identified
to come from the $W$ loop. We could  remark, in passing,
that the $\gamma \gamma \to \gamma \gamma $ amplitude at high energies
has exactly the structure anticipated long ago by combining
the Vector Meson Dominance (VMD) idea, with the assumption that the Pomeron
couplings are predominantly helicity-non-flip!  But, of course in the
present theory, the role of the Pomeron  is played by the $W$ loop,
and the aforementioned
success of VMD seems accidental! \par

As it has been recently emphasized in \cite{GPRgamma1},
this remarkable property suggests to use the
$\gamma \gamma \to \gamma \gamma$ scattering
process as a tool for searching for types of new physics characterized
by amplitudes with a substantial imaginary part; like \eg~ effects
due to chargino or charged slepton  loop diagrams above the threshold;
$s$-channel resonance production;
or new strong interactions inducing
unitarity saturating contributions to the NP amplitudes.\par

In the present paper  we study in detail the $\gamma \gamma \to
\gamma \gamma $ amplitudes in the standard and SUSY models. The
idea behind this, is to use $\gamma \gamma \to \gamma \gamma $
scattering for searching for SUSY signatures. The situation for
such a search should be particularly favorable at energies
above the  charged supersymmetric particle threshold, where  the SUSY
contribution to the $\gamma \gamma \to \gamma \gamma $ amplitude
has a large imaginary part interfering effectively  with the
standard one. Such a search is complementary to the direct
production of charged SUSY particles and it should help
identifying their nature; since it avoids the
model-dependent task of studying their decay modes, once they
are actually produced. More explicitly: the charged sparticle loop
contribution to $\gamma \gamma \to \gamma \gamma $, is independent
of the many  parameters entering their decay modes and determining
\eg~ the soft SUSY breaking and the possible $R$-parity violating
sectors. \par

The expressions for the $W$ and fermion loop
contributions are of course well known \cite{Jikia, vanderBij}, but
their detailed properties had not been fully analyzed before. We
have confirmed the results  of \cite{Jikia, vanderBij},
 using the non-linear gauge of
\cite{Dicus}, and we give  them in Appendix A,  together
with  the  1-loop contribution induced by  a single charged scalar
particle. The rest of the contents of the paper is the following:
In Section 2,  a simple and accurate high energy approximation  to
the SM $\gamma \gamma \to \gamma \gamma $ amplitude is presented,
which elucidates very clearly its physical properties in SM at
high energies, and
should be useful for identifying  certain forms of New
Physics (NP) contributing to it.  We consider SUSY, as
an example of such an NP,  and we discuss the physical properties of
the contribution to the above amplitudes from a chargino or charged
slepton; which may be expected to be lighter than $\sim 250$GeV
\cite{LC-PhysRep}. In Sec. 3, we study the $\gamma \gamma \to
\gamma \gamma $ cross sections in the standard and  SUSY models,
for various polarizations of the incoming photons. We identify the
sensitivity of these cross sections to various SUSY effects and we
discuss their observability in unpolarized and polarized $\gamma \gamma$
collisions, realized through the present ideas of laser backscattering.
In Appendix B we summarize the laser backscattering
formalism and give the expressions of the $\gamma \gamma$
flux and the two-photon spin density matrix \cite{LCgg}.
Finally, in Sec. 4, we summarize the results and
give our conclusions.

\section{An overall view of the $\gamma \gamma \to \gamma \gamma $
amplitudes.}

The invariant helicity amplitudes  $F_{\lambda_1 \lambda_2
\lambda_3\lambda_4}(\s,\t,\u)$ for the process $\gamma \gamma \to \gamma
\gamma $ are given in Appendix A. Altogether there are $2^4=16$
helicity amplitudes, which must of course satisfy the
constraints  from Bose (\ref{Bose1}, \ref{Bose2}) and crossing
symmetry (\ref{cross-ts}, \ref{cross-us}).
In SM or  SUSY models,
parity  and time inversion invariance also hold, which imply
(\ref{parity})) and (\ref{time-inv}) respectively, thereby
allowing to express all helicity amplitudes in terms of the analytic
expressions of just the three functions  $F_{+++-}(\s,\t,\u)$,
$F_{++--}(\s,\t,\u)$ and $F_{++++}(\s,\t,\u)$ \cite{Jikia}; compare
(\ref{+++-} - \ref{++++2}).
In Appendix A, we reproduce the $W$ and charged fermion  contributions
of  \cite{Jikia} and \cite{vanderBij}  respectively, and we also
give the charged scalar loop contributions to these amplitudes.\par

All results are given in terms of the standard 1-loop functions
$B_0$, $C_0$ and $D_0$, first introduced in \cite{Passarino}.
For the special   photon scattering case we are interested in,
these  functions may be written as  $B_0(s)$, $C_0(s)$,
$D_0(s,t)$, following the definitions in (\ref{B0} -
\ref{D0}). These functions depend only on the indicated variables and
the mass $m$ of the particle circulating in the loop.
In the SM case, the role of the mass $m$ is played by either
$W$ mass or the masses of the quarks and charged leptons.
This means that in the kinematical region
relevant for a Linear Collider, we  have
$\s, |\t|, |\u| \gg m^2$; apart of course from the $t$-quark case,
which  is
not very important for the overall magnitude.
It turns out that  for high    $(\s, |\t|, |\u|  )$,
an  excellent approximation to the above
1-loop functions is given by
\bqa
B_0(\s)& \simeq & \Delta +2 -Ln \left (\frac{-\s -i\epsilon}{\mu^2} \right )
\ ~ , \label{B0asym}\\
C_0(\s) & \simeq & \frac{1}{2 \s} \left [Ln \left (\frac{-\s -i\epsilon}{m^2}
\right ) \right ]^2 \ \ , \label{C0asym} \\
D_0(\s,\t) & \simeq & \frac{2}{\s \t}
\Bigg [ Ln \left (\frac{-\s -i\epsilon}{m^2}
\right ) Ln \left (\frac{-\t -i\epsilon}{m^2} \right )
-~\frac{\pi^2}{2} \Bigg ] \ \ \label{D0asym},
\eqa
where $\Delta $ is the usual infinite term entering the calculation
of  the  divergent integral
for $B_0(\s)$, and  $\mu$ is the dimensional regularization  parameter
  \cite{Hagiwara}.  These results  can be easily obtained
by keeping the leading term in a $m^2/\s, ~ m^2/\t$ expansion of the
formulae in the Appendix of \cite{Denner}. Numerically
they are extremely accurate, provided that  $|\s| \gsim 100~ m^2$ in
(\ref{B0asym}, \ref{C0asym}); while a similar accuracy
for (\ref{D0asym})
 obtains in the region
\bqa
\label{D0asym-region}
& -\s \gsim 100~ m^2 ~~~& ,   ~~~~
 -\t \gsim 100~ m^2 ~~~, \nonumber \\
& \mbox{or} &~~~~~  \s > -\t \gsim 100~ m^2 ~~~, ~~~~~
\nonumber \\
& \mbox{or} & ~~~~~  \t > -\s \gsim 100~ m^2  \ \ .
\eqa

We can now obtain simple expressions for the $W$ and light fermion
contributions to the $\gamma \gamma \to \gamma \gamma $ amplitudes,
which should be quite accurate for the large energies and scattering angles
relevant for  $LC_{\gamma \gamma}$ experiments.
Substituting thus, (\ref{B0asym} - \ref{D0asym}) in
(\ref{W++++} - \ref{W++--}) and neglecting all terms of order $\mwd/\s$,
$\mwd/\t$, $\mwd/\u$, we get
\bqa
 &&\frac{F^W_{++++} (\s,\t,\u)}{\alpha^2}  \simeq  12
+ 12 ~\left (\frac{\u-\t}{\s} \right )
\left \{ Ln \left (\frac{-\u -i\epsilon}{m^2} \right ) -
Ln \left (\frac{-\t -i\epsilon}{m^2} \right ) \right \}
\nonumber \\
&& +  16 \left( 1 -~\frac{3\t\u}{4\s^2} \right )
\Bigg [ \left \{ Ln \left (\frac{-\u -i\epsilon}{m^2} \right ) -
Ln \left (\frac{-\t -i\epsilon}{m^2} \right ) \right \}^2 +\pi^2
\Bigg ]
\nonumber \\
&& +  16 \s^2 \Bigg \{  \frac {1}{\s\t}\,
Ln \left (\frac{-\s -i\epsilon}{m^2}
\right ) Ln \left (\frac{-\t -i\epsilon}{m^2} \right ) +
 \frac {1}{\s\u}\, Ln \left (\frac{-\s -i\epsilon}{m^2}
\right ) Ln \left (\frac{-\u -i\epsilon}{m^2} \right )
\nonumber \\
&& + \frac {1}{\t\u}\, Ln \left (\frac{-\t -i\epsilon}{m^2}
\right ) Ln \left (\frac{-\u -i\epsilon}{m^2} \right ) \Bigg \}
~~ , \label{Wlarge}
\eqa
\bq
F^W_{+++-} (\s,\t,\u) \simeq F^W_{++--} (\s,\t,\u) \simeq -12 \alpha^2
\simeq \mbox{negligible} ~~ . \label{Wsmall}
\eq
Correspondingly, the asymptotic  expressions for a single fermion loop of
charge $Q_f$ and mass $m_f$, derived from
(\ref{f++++} - \ref{f++--}) by neglecting all terms of $O(m_f^2 /\s)$,
$O(m_f^2 /\t)$, $O(m_f^2 /\u)$, are
\bqa
 &&\frac{F^f_{++++} (\s,\t,\u)}{\alpha^2 Q_f^4}  \simeq  -8
- 8 ~\left (\frac{\u-\t}{\s} \right )
\left \{ Ln \left (\frac{-\u -i\epsilon}{m^2} \right ) -
Ln \left (\frac{-\t -i\epsilon}{m^2} \right ) \right \}
\nonumber \\
&& -  4 ~ \frac{(\t^2+ \u^2)}{\s^2}
\Bigg [ \left \{ Ln \left (\frac{-\u -i\epsilon}{m^2} \right ) -
Ln \left (\frac{-\t -i\epsilon}{m^2} \right ) \right \}^2 +\pi^2
\Bigg ] ~~ , \label{fasym++++}
\eqa
\bq
F^f_{+++-} (\s,\t,\u) \simeq F^f_{++--} (\s,\t,\u) \simeq 8 Q_f^4 \alpha^2
\simeq \mbox{negligible} ~~ . \label{fasym+++-}
\eq

On the basis of  (\ref{Wlarge} - \ref{fasym+++-})
and (\ref{+++-} - \ref{++++2}), we  see
that \underline{in the Standard Model},
the only physical amplitudes which have
a chance of being non-negligible
at LC energies, are  $F_{\pm\pm\pm\pm}(\s,\t,\u)$ and
$F_{\pm\mp\pm\mp}(\s,\t,\u)=F_{\pm\mp\mp\pm}(\s,\u, \t)$.
Indeed a detail look at the aforementioned equations shows
that these are the \underline{only} amplitudes which  (may generally)
receive a logarithmically enhanced high energy contribution.
In the physical region
of the scattering amplitudes, such a contribution  is almost purely
imaginary and arises   from the term within the last curly brackets
of the $W$ loop expression (\ref{Wlarge}).  \par

The real contributions to the various amplitudes are much
smaller. For the physical amplitude $F_{\pm\pm\pm\pm}$, the most
important real contribution  below $1~TeV$, arises from the last term
in (\ref{fasym++++}).  Its origin is fermionic
and it is enhanced not by a logarithm, but rather by
a  large $\pi^2$ term. In this energy range,
there exist also a somewhat  smaller real contribution
affecting the $F_{\pm\pm\pm\pm}(\s,\t,\u)$ and $
F_{\pm\mp\pm\mp}(\s,\t,\u)=F_{\pm\mp\mp\pm}(\s,\u, \t)$
amplitudes, which is due to some linear $log$ terms;
while the Sudakov-type  $log^2$  terms cancel out at
both, large ($\s \sim -\t/2 \sim -\u/2$) and small angles
($\s \gg -\t$ or $\s \gg -\u$. In any case, it should be noted,
that the real part of all the large amplitudes is
always more than five times smaller than the imaginary part.\par

Numerical results for these amplitudes using the exact 1-loop functions
have been presented in Fig.1 of \cite{GPRgamma1}, and they are quite
similar to the results obtained from (\ref{Wlarge} - \ref{fasym+++-}).
Concerning the accuracy of the above asymptotic expressions at
$LC_{\gamma \gamma}$ energies, we note that for the large amplitudes
cases of $F_{++++}$ and $F_{+-+-}=F_{+--+}$, the asymptotic expressions
tend to be higher than the exact 1-loop ones by $\sim 20\%$ at
about $0.4~ TeV$,
and by less than $10\%$ as we approach $1~TeV$. For the small
amplitudes cases,
the relative accuracy may occasionally be not so good, but this is
not relevant, since they are really negligible. To complete the
 discussion about the SM amplitudes, we also note that the top
contribution
is at least an order of magnitude smaller than the other SM
contributions we have just discussed. \par

The approximate SM amplitudes in (\ref{Wlarge} - \ref{fasym+++-})
can  then be used to understand the  magnitude of the NP
contribution to the $\gamma \gamma \to \gamma \gamma $ cross
sections, under various polarizations conditions. These suggest
that $\gamma \gamma \to \gamma \gamma $ scattering may  provide a
very useful tool for searching for types of New Physics (NP),
with largely  imaginary amplitudes \cite{GPRgamma1}. \par

\begin{figure}[p]
\vspace*{-4cm}
\[
\epsfig{file=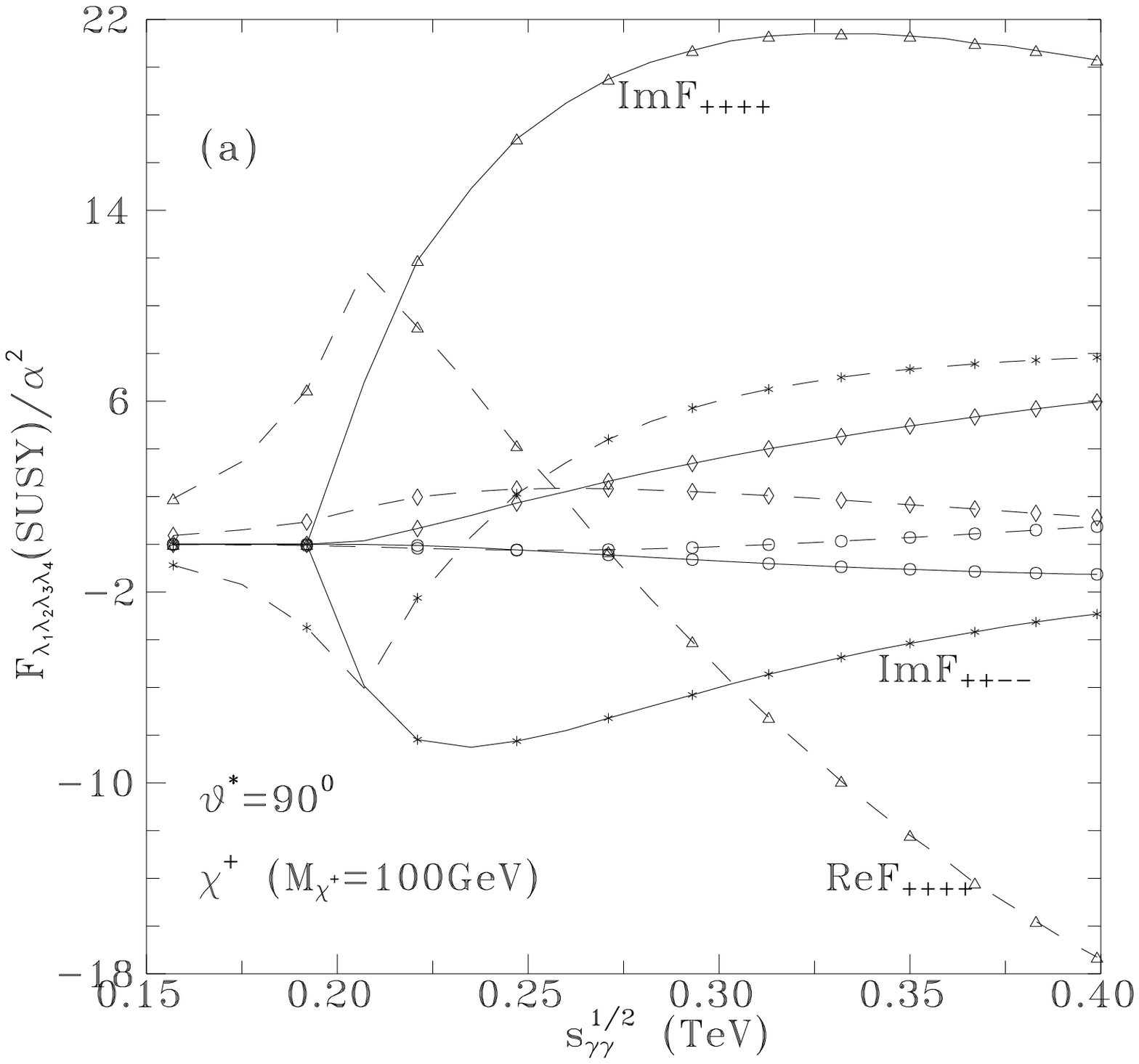,height=7.5cm}\hspace{0.5cm}
\epsfig{file=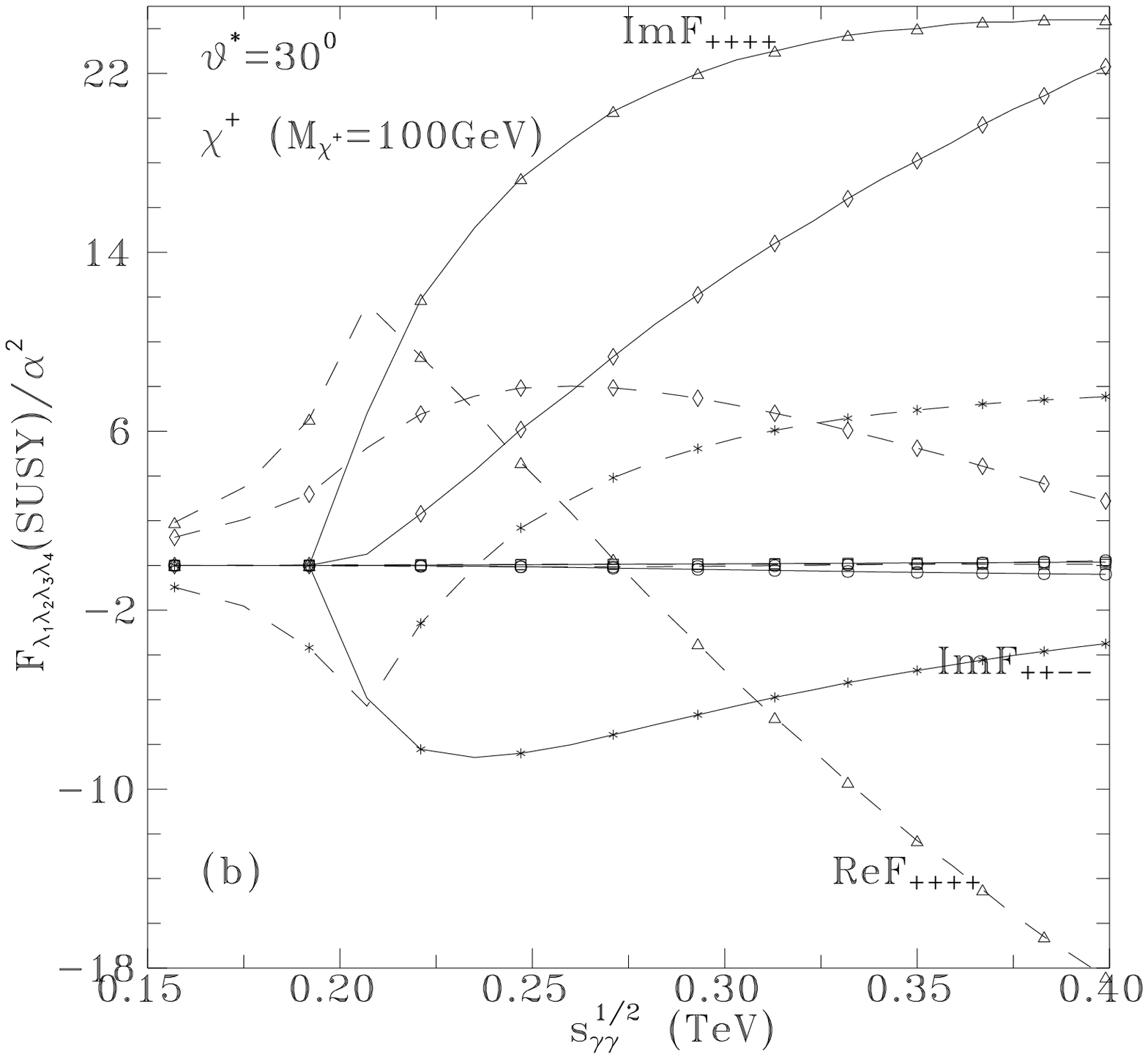,height=7.5cm}
\]
\vspace*{1.5cm}
\[
\epsfig{file=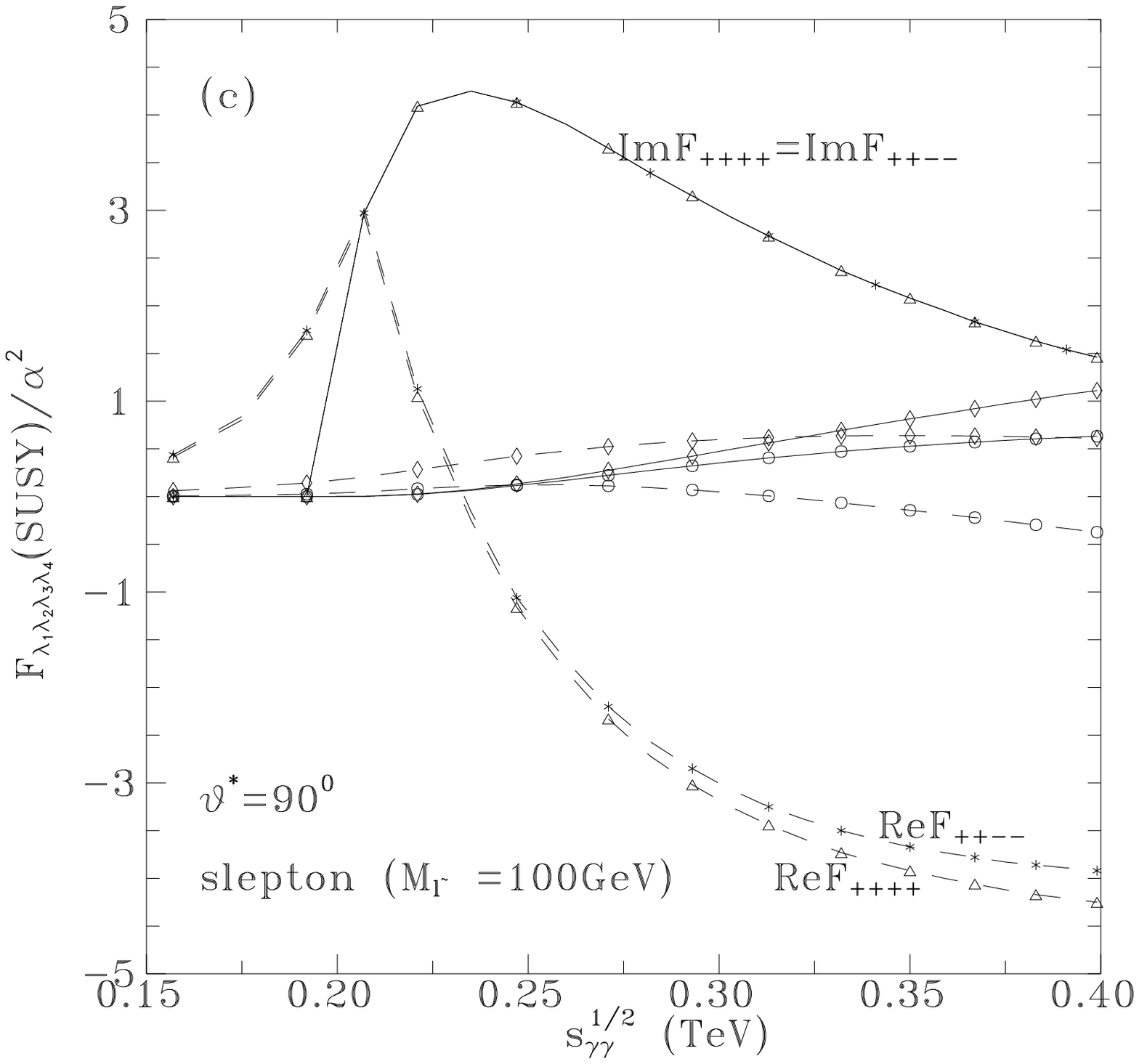,height=7.5cm}\hspace{0.5cm}
\epsfig{file=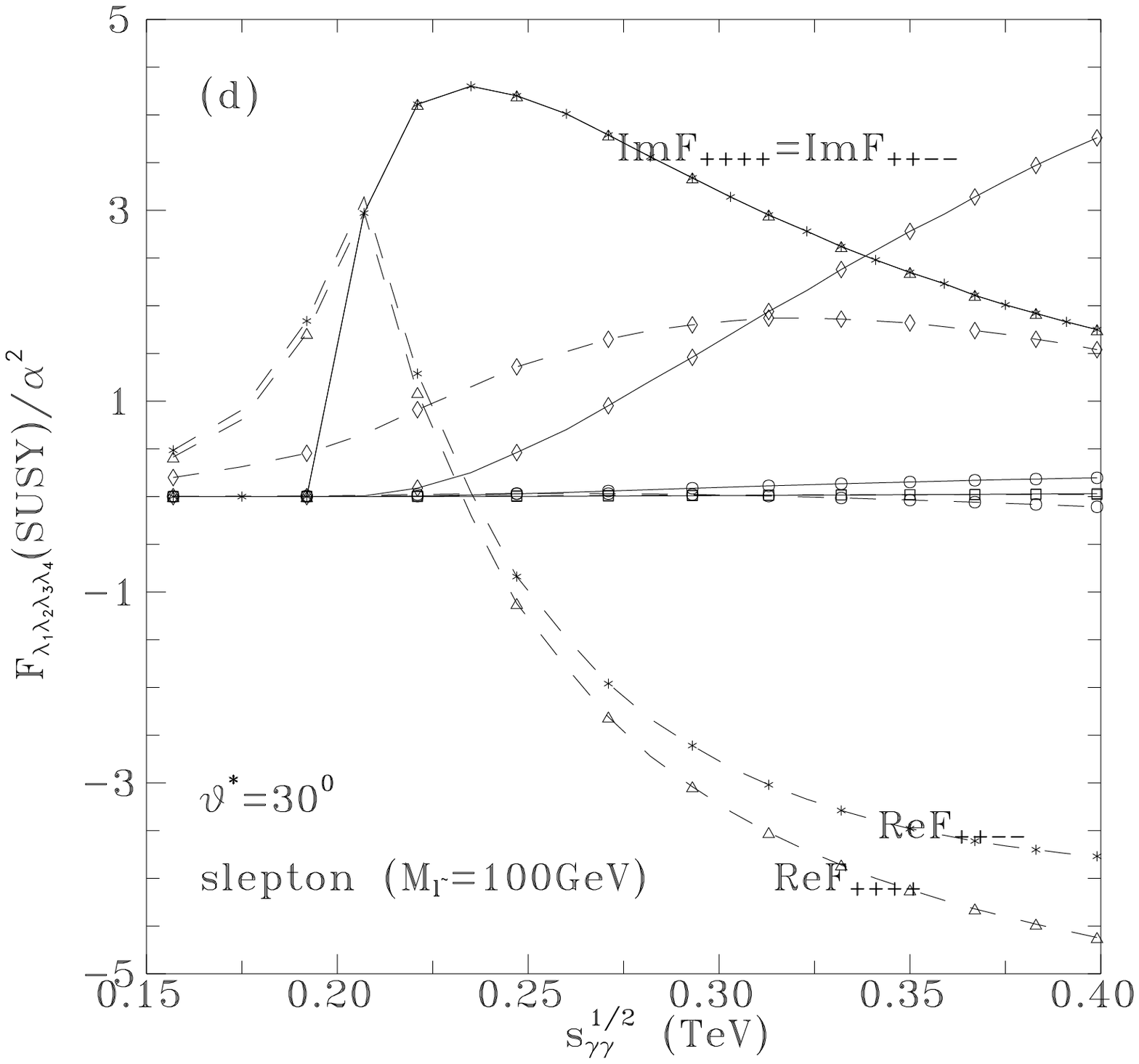,height=7.5cm}
\]
\vspace*{1.cm}
\caption[1]{Imaginary (solid) and real (dash) parts
of the chargino (a,b) and slepton (c,d)
contributions to the $\gamma \gamma \to \gamma \gamma
$ helicity amplitudes
at $\vartheta =90^0$ (a,c), and $\vartheta =30^0$ (b,d).
The notation is: $F_{++++}$ (triangles),
$F_{+++-}$ (circles), $F_{++--}$ (stars),
$F_{+-+-}$ (rhombs).
$F_{+--+}$, is identical to $F_{+-+-}$ for the (a,c) cases,
while it is given by 'boxes' in the (b,d) ones.}
\label{chargino-amp}
\end{figure}

Thus in Fig.\ref{chargino-amp}a,b we give the contributions from
a chargino of mass 100~GeV for two values of the c.m. scattering angle,
derived from (\ref{f++++} -\ref{f++--}), on the basis of the exact
expressions for the 1-loop functions \cite{Oldenborgh}.
The corresponding
results for a slepton, are derived using (\ref{slepton++++} -
\ref{slepton++--}) and presented in
Fig.\ref{chargino-amp}c,d. As seen in both cases,
immediately  above the threshold, a considerable imaginary
contribution to the $F_{++++}$ amplitude starts developing,
which can interfere with the SM one and produce a measurable effect.
We also note, that the slepton contribution is considerably
smaller than the chargino one, but, as we will see below,
the effect may increase if several scalar
sparticles (charged sleptons, $\tilde t_1$ or $\tilde H^+$) appear
below $250GeV$. \par

\section{ The $\gamma \gamma \to \gamma \gamma $ Cross sections}

We next explore   the possibility to use polarized or unpolarized
$\gamma\gamma$ collisions in an LC operated in the $\gamma
\gamma $ mode,  through laser backscattering
\cite{Tsi, GPRgamma1}.  Bose statistics and the assumption of
Parity invariance leads to the following form for the
$\gamma \gamma \to \gamma \gamma $
cross section
\bqa
{d\sigma\over d\tau d\cos\vartheta^*}&=&{d \bar L_{\gamma\gamma}\over
d\tau} \Bigg \{
{d\bar{\sigma}_0\over d\cos\vartheta^*}
+\langle \xi_2 \xi_2^\prime \rangle{d\bar{\sigma}_{22}\over d\cos\vartheta^*}
+[\langle\xi_3\rangle\cos2\phi+\langle\xi_3^ \prime\rangle\cos2\phi^\prime]
{d\bar{\sigma}_{3}\over d\cos\vartheta^*}
\nonumber\\
&&+\langle\xi_3 \xi_3^\prime\rangle[{d\bar{\sigma}_{33}\over d\cos\vartheta^*}
\cos2(\phi+\phi^\prime)
+{d\bar{\sigma}^\prime_{33}\over
d\cos\vartheta^*}\cos2(\phi- \phi^\prime)]\nonumber\\
&&+[\langle\xi_2 \xi_3^\prime\rangle\sin2 \phi^\prime-
\langle\xi_3 \xi_2^\prime\rangle\sin2\phi]
{d\bar{\sigma}_{23}\over d\cos\vartheta^*} \Bigg \} \ \ ,
\label{sigpol}
\eqa

where
\bqa
{d\bar \sigma_0\over d\cos\vartheta^*}&=&
\left ({1\over128\pi\hat{s}}\right )
\sum_{\lambda_3\lambda_4} [|F_{++\lambda_3\lambda_4}|^2
+|F_{+-\lambda_3\lambda_4}|^2] ~ ,  \label{sig0} \\
{d\bar{\sigma}_{22}\over d\cos\vartheta^*} &=&
\left ({1\over128\pi\hat{s}}\right )\sum_{\lambda_3\lambda_4}
[|F_{++\lambda_3\lambda_4}|^2
-|F_{+-\lambda_3\lambda_4}|^2]  \ , \label{sig22} \\
{d\bar{\sigma}_{3}\over d\cos\vartheta^*} &=&
\left ({-1\over64\pi\hat{s}}\right ) \sum_{\lambda_3\lambda_4}
Re[F_{++\lambda_3\lambda_4}F^*_{-+\lambda_3\lambda_4}]  \ ,
\label{sig3} \\
{d\bar \sigma_{33} \over d\cos\vartheta^*}& = &
\left ({1\over128\pi\hat{s}}\right ) \sum_{\lambda_3\lambda_4}
Re[F_{+-\lambda_3\lambda_4}F^*_{-+\lambda_3\lambda_4}] \ ,
\label{sig33} \\
{d\bar{\sigma}^\prime_{33}\over d\cos\vartheta^*} &=&
\left ({1\over128\pi\hat{s}}\right ) \sum_{\lambda_3\lambda_4}
Re[F_{++\lambda_3\lambda_4}F^*_{--\lambda_3\lambda_4}] \  ,
\label{sig33prime} \\
{d\bar{\sigma}_{23}\over d\cos\vartheta^*}& = &
\left ({1\over64\pi\hat{s}}\right ) \sum_{\lambda_3\lambda_4}
Im[F_{++\lambda_3\lambda_4}F^*_{+-\lambda_3\lambda_4}] \ ,
\label{sig23}
\eqa
are expressed in terms of the $\gamma \gamma \to \gamma \gamma$
amplitudes given in
Appendix A. Note that only $d\bar \sigma_0/ d\cos\vartheta^*$
is positive definite.\par

  The quantity $d\bar L_{\gamma\gamma}/d\tau$
 (compare (\ref{sigpol}), (\ref{Lgammagamma})) describes the
photon-photon luminosity
per unit $e^-e^+$ flux, in an LC operated in the $\gamma \gamma$ mode
\cite{LCgg}. Moreover, $\vartheta^*$ is the scattering angle in
the $\gamma \gamma $ rest frame and
$\tau \equiv s_{\gamma \gamma}/s_{ee}$.
The Stokes parameters $\xi_2$, $\xi_3$ and the azimuthal angle
$\phi$ in (\ref{sigpol}), determine the normalized helicity density matrix
of one of the backscattered photons $\rho^{BN}_{\lambda \lambda^\prime}$
through the formalism in Appendix B; compare (\ref{rhoBN}) \cite{Tsi}.
The corresponding parameters for the other backscattered photon are
denoted by a prime. \par

\begin{figure}[p]
\vspace*{-3cm}
\[
\epsfig{file=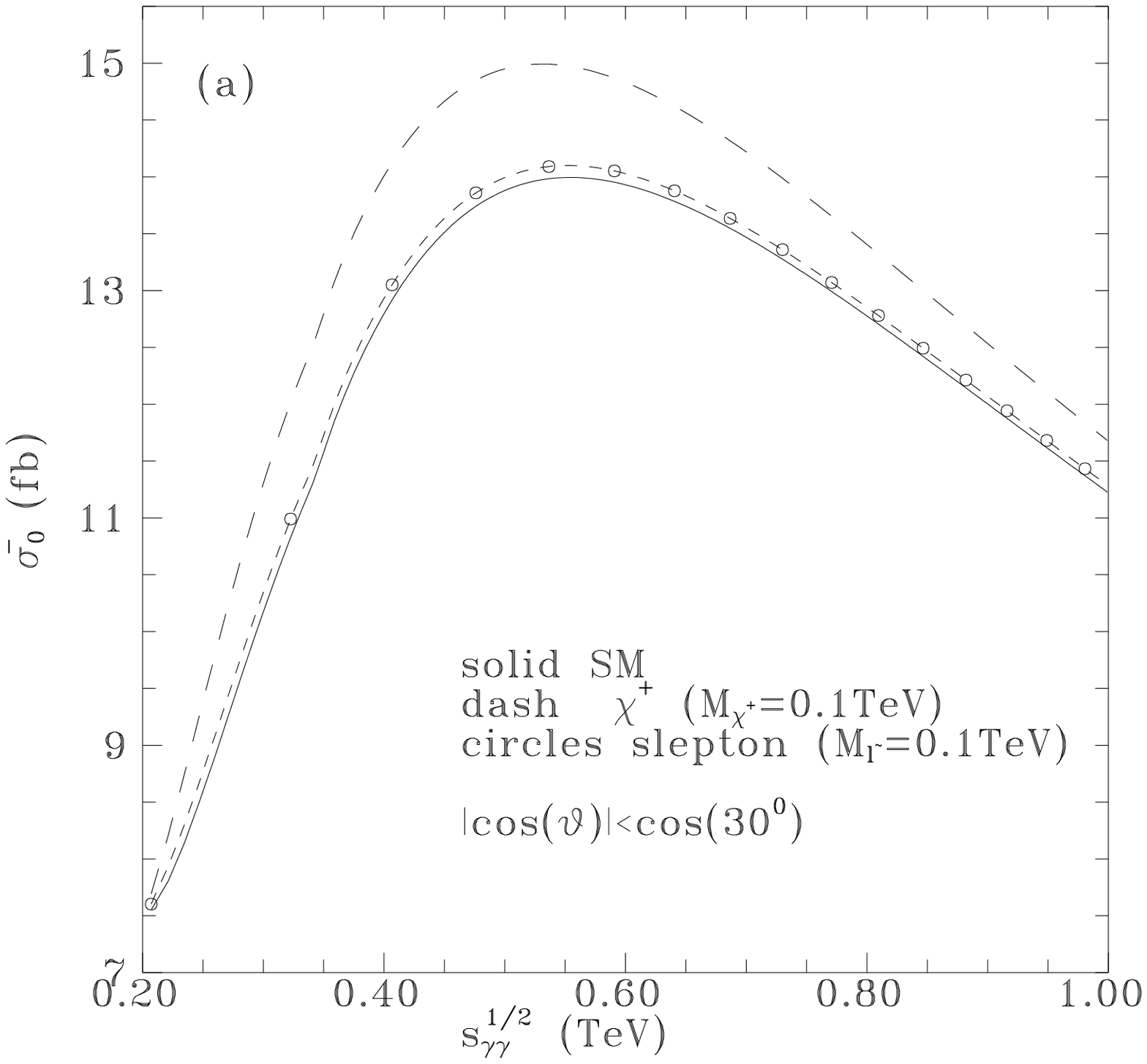,height=7.5cm}\hspace{0.5cm}
\epsfig{file=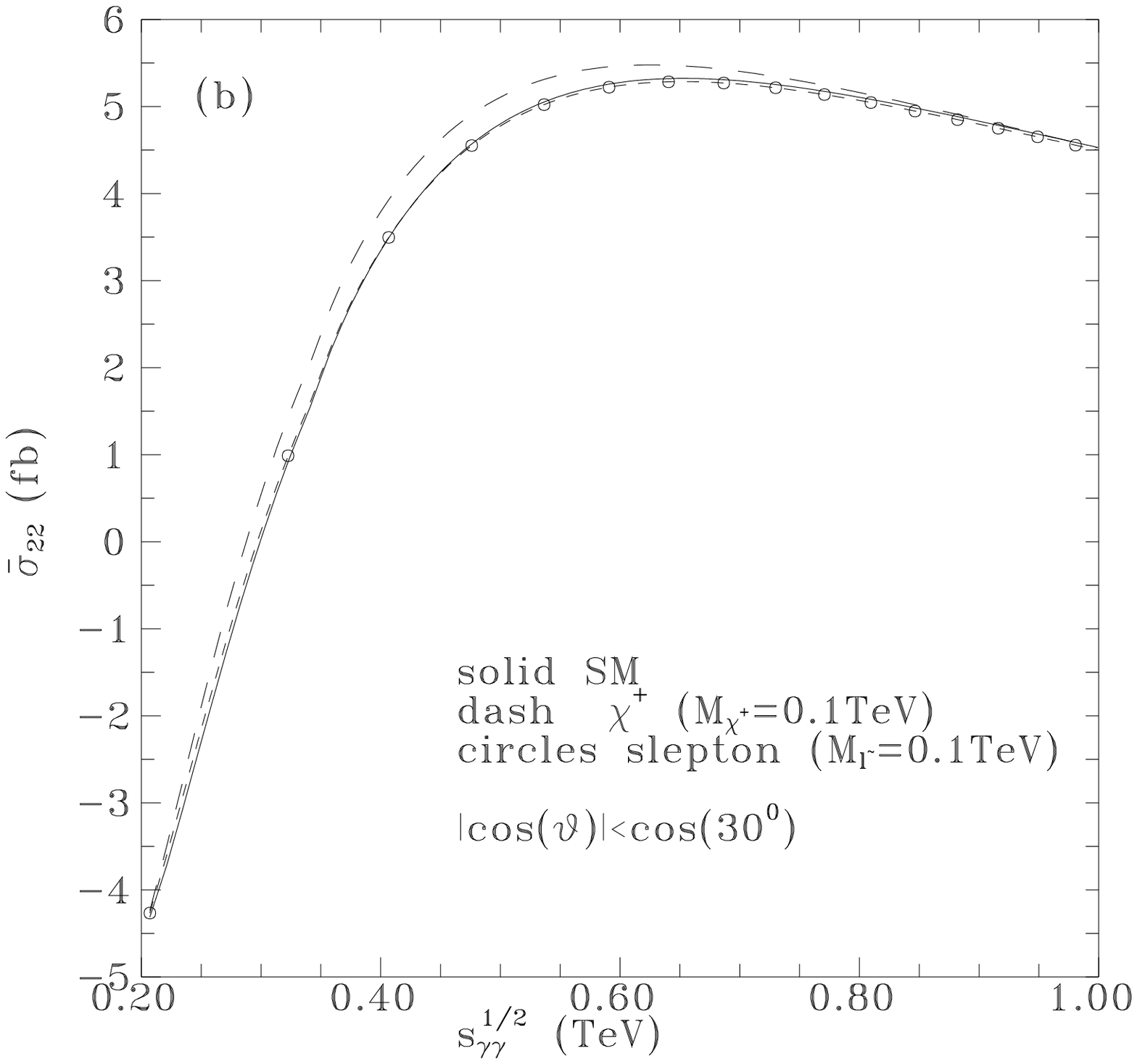,height=7.5cm}
\]
\vspace*{1.5cm}
\[
\epsfig{file=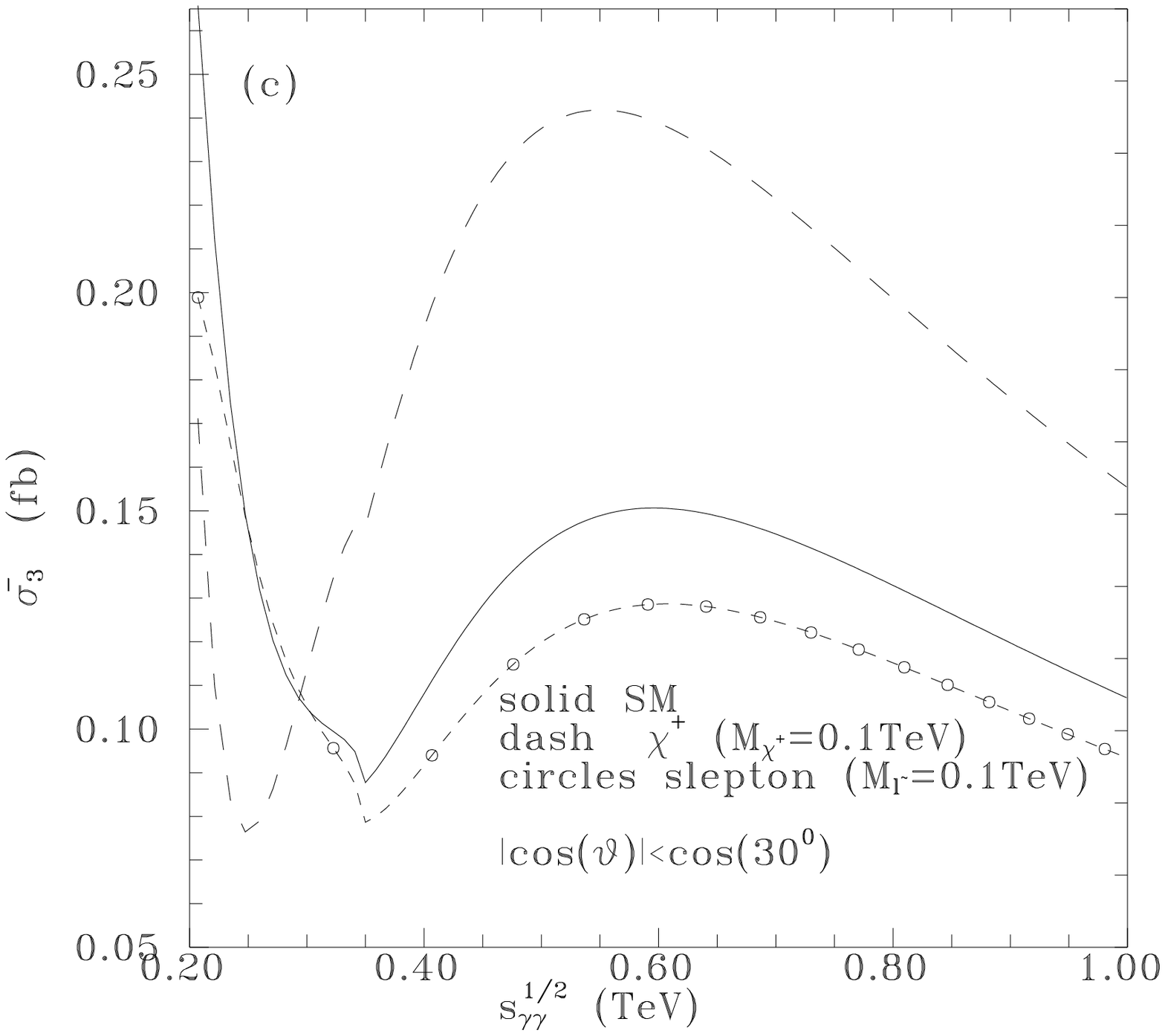,height=7.5cm}\hspace{0.5cm}
\epsfig{file=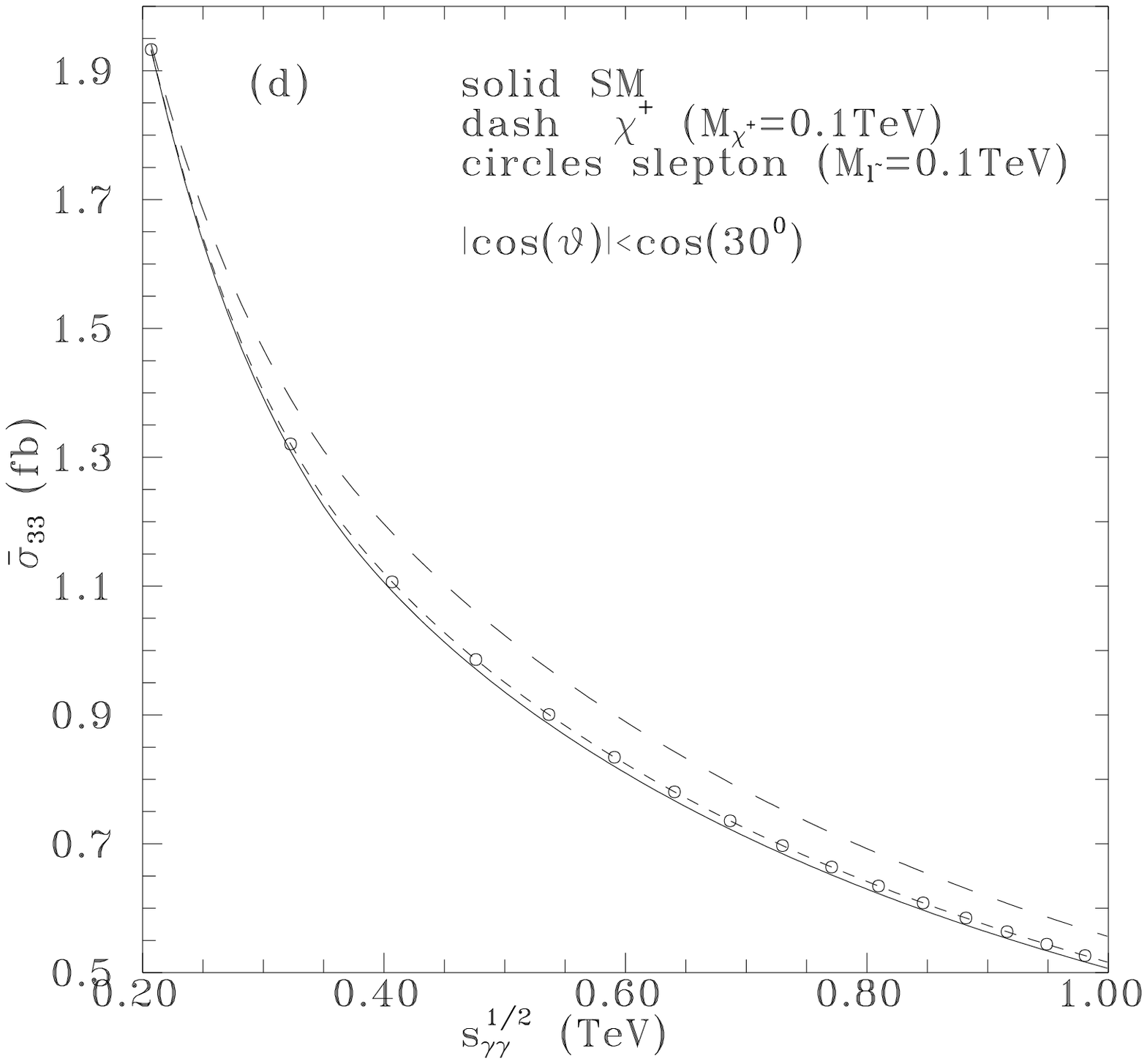,height=7.5cm}
\]
\vspace*{1.cm}
\caption[1]{$\bar \sigma_0$, $\bar \sigma_{22}$, $\bar \sigma_3 $
and  $\bar \sigma_{33} $ for SM (solid) and in the presence
of a chargino (dash) or a charged slepton (circles) contribution. }
\label{SUSY-sig0-33}
\end{figure}

The results for the cross sections $\bar \sigma_j$, integrated in the
range $30^0 \leq \vartheta^* \leq 150^0$,  are given in
Fig.\ref{SUSY-sig0-33}a-f, for the standard model,
as well as for the   case  including the contributions from
a single chargino or a single charged slepton with mass 100 GeV.
In Fig.\ref{SUSY-sig0-33-2}a-f the corresponding results for
a 250 GeV SUSY mass are given.
Note that the charged slepton results will also be valid for the charged
Higgs case; while for  a single $\tilde t$ contribution, the SUSY effect
will be reduced by a factor $3Q^4_{\tilde t}= 3 (2/3)^4 \simeq 0.59$.
As seen from Fig.\ref{SUSY-sig0-33}a-f,\ref{SUSY-sig0-33-2}a-f,
 the chargino and slepton contributions to
$\bar \sigma_3$ and  $\bar \sigma_{33}^\prime$ are mostly of
opposite sign; as opposed to the $\bar \sigma_0$, $\bar \sigma_{22}$
and $\bar \sigma_{33}$ cases where the signs are usually the same.
For $\bar \sigma_{23}$ an intermediate situation appears, in which
 the chargino and slepton
contributions tend to be of opposite sign for
$M_\chi \sim M_{\tilde l} \sim 100 GeV$, but they are mostly of
the same sign if $M_\chi \sim M_{\tilde l} \sim 250 GeV$; compare
Fig.\ref{SUSY-sig33p-23}f,\ref{SUSY-sig33p-23-2}f.\par

Unfortunately, as seen from Fig.\ref{SUSY-sig0-33}c,e and
Fig.\ref{SUSY-sig0-33-2}c,e, the
quantities $\bar \sigma_3$ and $\bar \sigma_{33}^\prime$, which
are most sensitive to the nature of the contributing sparticles,
are numerically the smallest ones. For studying therefore
SUSY-type NP, we have to rely mainly on the largest  quantity
$\bar \sigma_0$ appearing in Fig.\ref{SUSY-sig0-33}a. Depending on
the experimental situation though,  $\bar \sigma_{22}$ given in
Fig.\ref{SUSY-sig0-33}b, should also prove useful. This, of course,
should not lead us to the idea that those $\bar \sigma_j$, which
are small in SM and SUSY, are not interesting; since there may
exist other forms of NP for which they
are sizable. It would therefore be important to study them and
bound their magnitude, in order to check at least the consistency with
SM and/or  SUSY.\par

To get a feeling of the observability  of the various quantities
$\bar \sigma_j$ appearing in (\ref{sigpol}), we
next turn to the experimental aspects of the $\gamma \gamma$ collision
process realized through the laser backscattering \cite{LC, LCgg}.
The general form of the overall  luminosity
$d\bar L_{\gamma\gamma}/d\tau$ and of the density matrix of the photon
pair, are  given in Appendix B; based on  the assumption
that the conversion point where the
Compton backscattering occurs, coincides with the interaction point
at which the $\gamma \gamma $
collision takes place \cite{LCgg}. It should be noticed  that
$d\bar L_{\gamma\gamma}/ d\tau$ depends on the frequencies,
of the two lasers, through  the parameters $x_0$ and
$x_0^\prime$ of (\ref{laser-kin}); and on the product of
longitudinal $e^\pm$ and laser polarizations
$P_e P_\gamma$ and $P_e^\prime P_\gamma^\prime$. As a result,
$d\bar L_{\gamma\gamma}/ d\tau$ becomes harder
as $P_e P_\gamma \to -1$,  or as $x_0$  or $x_0^\prime$
approach their maximum value $ 2 (1+\sqrt{2})$;
(compare Fig.\ref{flux-dis}).\par

For obtaining the number of the expected events in each case, the
cross sections in (\ref{sigpol}) should be multiplied by the
$e^+e^-$ luminosity $\L_{ee}$, whose  presently contemplated value
for the LC project is $\L_{ee} \simeq 500~-~1000~
fb^{-1}$ per one or two years of running in \eg\@ the high
luminosity TESLA mode at energies of $350-800~ GeV$  \cite{LC}.\par

To this aim, we first express  $\bar \sigma_j$, multiplied
by their $\gamma \gamma$ luminosity coefficient in (\ref{sigpol}),
in terms  of  linear combinations of cross sections for various
longitudinal and/or transverse polarizations of the $e^\pm$ and laser beams.
Thus, for unpolarized $e^\pm$ and laser beams,
$\bar \sigma_0$ can be measured through
\bq
\left (\frac{d\bar L_{\gamma\gamma}}{ d\tau}\right )
{d\bar{\sigma}_{0}\over d\cos\vartheta^*}
= {d\sigma \over d\tau d\cos\vartheta^*}\Bigg \vert_{\mbox{unpol}}
~~~~ \ . \label{sigma0-unp}
\eq
On the other hand, by considering collisions with the combinations
of longitudinal polarizations
$(P_e, P_\gamma, P_e^\prime, P_\gamma^\prime)$ and
$(P_e, P_\gamma,  -P_e^\prime, -P_\gamma^\prime)$ and no transverse
polarizations, the quantities $\bar \sigma_0$ and $\bar \sigma_{22}$
can be measured through
\bqa
&&\left ({d\bar L_{\gamma\gamma}\over d\tau}\right )
{d\bar{\sigma}_{0}\over d\cos\vartheta^*}
 = \frac{1}{2}\Bigg [{d\sigma(P_e, P_\gamma,
P_e^\prime, P_\gamma^\prime ) \over d\tau d\cos\vartheta^*} +
 {d\sigma(P_e, P_\gamma, -P_e^\prime, -P_\gamma^\prime )
\over d\tau d\cos\vartheta^*} \Bigg ] \ , \label{sigma0-lon} \\
&&\left ({d\bar L_{\gamma\gamma}\over d\tau}\right )
\langle\xi_2 \xi_2^\prime \rangle
{d\bar{\sigma}_{22}\over d\cos\vartheta^*}
 = \frac{1}{2}\Bigg [{d\sigma(P_e, P_\gamma,
P_e^\prime, P_\gamma^\prime ) \over d\tau d\cos\vartheta^*} -
 {d\sigma (P_e, P_\gamma, -P_e^\prime, -P_\gamma^\prime )
\over d\tau d\cos\vartheta^*} \Bigg ] \ . \label{sigma22-lon}
\eqa
The results of (\ref{sigma0-unp} - \ref{sigma22-lon}) integrated
in the region $30^0 \leq \vartheta^* \leq 150^0$,  for the
indicated polarizations and the laser parameters
 $x_0=x_0^\prime=4.83$,
are presented in Fig.\ref{SUSY-sig-flux1} for a $100~GeV$ chargino
or slepton. \par

The measurement of $\bar \sigma_3$ could be achieved by selecting
one of the two laser photons to be purely transversely polarized
with \eg~ $P_t=1$ and direction determined by the azimuthal angle
$\phi$, while the other laser photon is taken unpolarized. In this
case $\bar \sigma_3$, together with $\bar \sigma_0$,  may be determined
through
\bqa
2\langle\xi_3\rangle \,\left({d\bar L_{\gamma\gamma}\over d\tau}\right )
{d\bar{\sigma}_{3}\over d\cos\vartheta^*} &= & {d\sigma(\phi=0) \over
d\tau d\cos\vartheta^*}~ -~  {d\sigma(\phi=\pi/2) \over d\tau
d\cos\vartheta^*} \ , \label{sigma3-t} \\
2 \, \left({d\bar L_{\gamma\gamma}\over d\tau}\right )
{d\bar{\sigma}_{0}\over d\cos\vartheta^*} &= & {d\sigma(\phi=0) \over
d\tau d\cos\vartheta^*}~ +~  {d\sigma(\phi=\pi/2) \over d\tau
d\cos\vartheta^*} \ . \label{sigma0-t}
\eqa
If  both laser photons are purely transversely polarized,
with $P_t=P_t^\prime=1$ and their  directions  determined by
the respective  azimuthal
angles $\phi,~\phi^\prime$; then $\bar \sigma_3$, $\bar
\sigma_{33}$, $\bar \sigma_{33}^\prime$, together with $\bar \sigma_0$
can  be determined through\footnote{Note that $\xi_3=\xi_3^\prime$
in this case.}
\bqa
2 \left({d \bar L_{\gamma\gamma}\over d\tau}\right )
{d\bar{\sigma}_{0}\over d\cos\vartheta^*}& =&
 {d\sigma(\phi=0, \phi^\prime= \pi/4) \over
d\tau d\cos\vartheta^*} + {d\sigma(\phi=\pi/4, \phi^\prime= \pi/2)
\over d\tau d\cos\vartheta^*} \ ,  \label{sigma0-tt} \\
 (\langle\xi_3\rangle+\langle\xi_3^\prime\rangle) \left({d \bar
L_{\gamma\gamma}\over d\tau}\right ) {d\bar{\sigma}_{3}\over
d\cos\vartheta^*}& =& {d\sigma(\phi=0, \phi^\prime= \pi/4) \over
d\tau d\cos\vartheta^*} - {d\sigma(\phi=\pi/4, \phi^\prime= \pi/2)
\over d\tau d\cos\vartheta^*} \ ,  \label{sigma3-tt} \\
2 \langle\xi_3 \xi_3^\prime\rangle
 \left({d\bar L_{\gamma\gamma}\over d\tau}\right )
{d\bar{\sigma}_{33}\over d\cos\vartheta^*}& = &{d\sigma(\phi=0 ,
\phi^\prime=0) \over d\tau d\cos\vartheta^*} -
{d\sigma(\phi=0, \phi^\prime= \pi/4) \over d\tau
d\cos\vartheta^*} \nonumber \\
&-& {d\sigma(\phi=\pi/4 , \phi^\prime=0) \over
d\tau d\cos\vartheta^*} + {d\sigma(\phi=\pi/4, \phi^\prime=- \pi/4)
\over d\tau d\cos\vartheta^*}  , \label{sigma33-tt} \\
2 \langle\xi_3 \xi_3^\prime\rangle
\left ({d\bar L_{\gamma\gamma}\over d\tau} \right )
{d\bar{\sigma}_{33}^\prime \over d\cos\vartheta^*}& = &{d\sigma(\phi=0 ,
\phi^\prime=0) \over d\tau d\cos\vartheta^*}
 - {d\sigma(\phi=0, \phi^\prime= \pi/4) \over d\tau
d\cos\vartheta^*} \nonumber \\
&-& {d\sigma(\phi=\pi/4 , \phi^\prime=0) \over
d\tau d\cos\vartheta^*} \
+ {d\sigma(\phi=\pi/4, \phi^\prime=+ \pi/4)
\over d\tau d\cos\vartheta^*} . \label{sigma33p-tt}
\eqa
The results of (\ref{sigma0-tt} - \ref{sigma33p-tt}), integrated in
the region $30^0 \leq \vartheta^* \leq 150^0$,  for the indicated
polarizations and SUSY masses are presented in
Fig.\ref{SUSY-sig-flux2}. In order
to increase sensitivity\footnote{ Which means increasing
$\xi_3,~ \xi_3^\prime$} as much as possible, we have
chosen $x_0=x_0^\prime=1$, which has the side effect of making the
$\gamma \gamma$ spectrum softer; (compare Fig.\ref{flux-dis}).

Finally, for studying $\bar \sigma_{23}$, we need a mixed
polarization situation, where one laser photon is longitudinally
polarized, while the other is transverse; like \eg\@
($P_e=0.8$, $P_\gamma =-1$, $P_t=0$) for the one, and
($P_e^\prime=P_\gamma^\prime =0$, $P_t^\prime=1$
with direction defined by $\phi^\prime$) for the other.
To optimize
the flux spectrum $d\bar L_{\gamma\gamma}/d\tau $, it may be better to
choose $x_0 \not= x_0^\prime$ in this case. In such case we have
\bqa
 2 \left ({d\bar L_{\gamma\gamma}\over d\tau}\right )
{d\bar{\sigma}_{0}\over
d\cos\vartheta^*}& =& {d\sigma(\phi^\prime= \pi/4) \over
d\tau d\cos\vartheta^*} + {d\sigma(\phi^\prime= 3\pi/4)
\over d\tau d\cos\vartheta^*} \ ,  \label{sigma0-lon-t} \\
 2 \langle\xi_2 \xi_3^\prime\rangle \left ({d \bar
L_{\gamma\gamma}\over d\tau}\right ) {d\bar{\sigma}_{23}\over
d\cos\vartheta^*}& =& {d\sigma(\phi^\prime= \pi/4) \over
d\tau d\cos\vartheta^*} - {d\sigma(\phi^\prime= 3\pi/4)
\over d\tau d\cos\vartheta^*} \ ,  \label{sigma23-lon-t} \\
\eqa
and an example appears in Fig.\ref{SUSY-sig-flux3}.
In this figure we also give predictions for an alternative
measurement of $\bar \sigma_3$; compare Fig.\ref{SUSY-sig-flux2}b
and Fig.\ref{SUSY-sig-flux3}b.\par

Using $\L_{ee}=500 fb^{-1}$, then  the $100~GeV$ chargino
effect indicated in
Fig.\ref{SUSY-sig-flux1}a for a $350~GeV$ LC and unpolarized $e^\pm$
and laser beams, is at the 2.3 standard deviations (SD) level;
while for the situation at Fig.\ref{SUSY-sig-flux1}b, it
increases 2.9 SD. In both cases, the effect arises from a
$\bar \sigma_0$ measurement, which itself measures the unpolarized
cross section. Nevertheless though, the sensitivity, as expressed
by the number of SD, does depend on the polarizations and $x_0$
parameters, since these affect the $\gamma \gamma$ flux  through
$d\bar L_{\gamma \gamma}/d\tau$; (compare (\ref{Lgammagamma})).
For studying therefore a suspected (due to some other signals)
chargino of a certain mass,
through $\gamma \gamma \to \gamma \gamma $ scattering, it
will be important to optimize the LC  and laser energies and $x_0$
parameters. To further elucidate this, we remark that for the
situations in Fig.\ref{SUSY-sig-flux2}a and
Fig.\ref{SUSY-sig-flux3}a, the chargino sensitivity
is at the 3.9 SD and 4.2 SD respectively. In all cases,
the $\tau$ regions used in estimating SD, are those
employed in the corresponding figures. \par

For the same $100~GeV$ chargino as above, the $\bar \sigma_{22}$
effect in Fig.\ref{SUSY-sig-flux1}c is at the 0.8 SD level, when a
bin like $0.49 \leq \tau \leq 0.62$ is used. Thus, a
$\bar \sigma_{22}$ measurement, which necessitates linear polarization,
can give an additional constraint.  \par

The quantities $\bar \sigma_{3}$ $\bar \sigma_{33}$ $\bar
\sigma_{33}^\prime$ and  $\bar \sigma_{23}$ are too small to be
measured with the above $\gamma \gamma $ flux, and the best we can
hope for, is to put some reasonable bound on them, which could help
excluding possible extreme forms of NP. \par

An analysis of the statistics of a $\bar \sigma_0$ measurement
for a $150~GeV$ and $250~GeV$ chargino was also made, and in these cases
we found that sensitivities at the 3 SD and 1.2 SD level should be
respectively expected. \par

As an example of the charged scalar case within the loop, we
considered the case of a single charged slepton.
If its mass is in the $100~GeV$ range, then the results in
Fig.\ref{SUSY-sig-flux1}a,b, \ref{SUSY-sig-flux2}a,
\ref{SUSY-sig-flux3}a would indicate a signal at the (0.5-0.7)SD
level.\par

The situation
may improve considerably though, if several, or even all six charged
sleptons expected in the minimal SUSY model, and maybe also
the lightest stop $\tilde t_1$ together with
one chargino, lie in (100-250) GeV mass region \cite{LC-PhysRep}.
A clearly measurable increase
(compared to the SM prediction),
may then appear in an $\bar \sigma_0$ measurement.
This is concluded from Figs.\ref{SUSY-sig0-33}a,
 \ref{SUSY-sig0-33-2}a, which show that in the (100-250)GeV mass
range, a fermion and scalar charged particle loop contribute with
the same sign to $\bar \sigma_0$.\par

\section{Conclusions}

In this paper we have offered a detailed analysis of the
helicity amplitudes of the process $\gamma \gamma \to \gamma
\gamma$ at high energies, and studied also the unpolarized
and polarized cross section.

The spectacular property of the Standard Model prediction for
this process is that, for energies above $0.3~TeV$,
there only two independent helicity amplitudes which are important;
namely $F_{\pm\pm\pm\pm}(\s,\t,\u)$ and
$F_{\pm\mp\pm\mp}(\s,\t,\u)=F_{\pm\mp\mp\pm}(\s,\u,\t)$.
These amplitudes are helicity conserving and almost purely imaginary for
all scattering angles. This property makes the $\gamma \gamma \to
\gamma \gamma$ process an excellent tool for searching for types
of new physics inducing large imaginary parts to such
amplitudes.\par

As such, we have studied here the particular SUSY case of a
single chargino or charged slepton contribution, at energies
above the threshold for their actual production. These
contributions depend of course only on the mass, charge and spin of
the SUSY partners, and are independent of the many model-dependent
parameters entering their decay modes. Thus, the study of the
$\gamma \gamma \to \gamma \gamma$ cross sections should offer
complementary  information, to the one obtained from direct
SUSY production cross sections.

For an LC collider at energies of $(350-800)GeV$ and a luminosity
$\L_{ee}=500fb^{-1}$,  using the presently contemplated ideas
about employing laser backscattering for transforming an LC to
$\gamma \gamma $ Collider,  we have found that the unpolarized
$\gamma \gamma \to \gamma \gamma$ cross section $\bar \sigma_0$, is
most sensitive to a chargino  loop contribution. In such a case, the
signal varies between a 3 SD and 1 SD effect,
as the chargino mass increases from $100$ to $250~GeV$. For a single
charged slepton with a $100~GeV$ mass, we have found that the
corresponding effect on $\bar
\sigma_0$ is at the $\sim$ (0.5-0.7)SD  level.\par

It is important to notice though, that in the $(100-250)GeV$ mass range,
both, the charged fermion and the
charged scalar particle loops, {\it increase } the SM
prediction for $\bar \sigma_0$. Thus, in the high energy limit,
this cross section gives a kind of counting of the number of states
involved in the loop.
Because of this and if  SUSY is realized in Nature below the
TeV-scale, then it would be quite plausible  that a
chargino, as well as all six charged sleptons  and $\tilde
t_1$, lie in the (100-250)GeV  mass range.
  In such a case, a clear signal could be seen
in $\bar \sigma _0$.\par

 The polarization quantities $\bar \sigma _3$
 or $\bar \sigma _{33}^\prime$, could in principle be used
to test the spin structure of the particles in the loop. However
with the foreseen photon-photon fluxes they are hardly
observable. Nevertheless, as fermion and scalar
loop contributions have different signs and tend to cancel
in these quantities, the exclusion of any effect would constitute
a valuable test of the global picture.\par

In any case it appears to us that the
$\gamma \gamma \to \gamma \gamma $ is a very clean process which
should supply an excellent tool for NP searches.
Further help, could also come from
corresponding effects in the $\gamma \gamma \to Z\gamma$ and
$\gamma \gamma \to ZZ$  processes, on
which we have already started working.
We conclude therefore, that important physical information
could arise from the study of  the
$\gamma \gamma \to \gamma \gamma$ process, and that his
 certainly constitutes
an  argument favoring the availability of the
laser $\gamma \gamma $ option in a Linear Collider.

\newpage
\renewcommand{\theequation}{A.\arabic{equation}}
\renewcommand{\thesection}{A.\arabic{section}}
\setcounter{equation}{0}
\setcounter{section}{0}

{\large \bf Appendix A: The $\gamma \gamma \to \gamma \gamma $
amplitudes in SM and SUSY.}

The invariant helicity amplitudes
 for  the process
\bq \gamma (p_1,\lambda_1) \gamma (p_2,\lambda_2) \to \gamma
(p_3,\lambda_3) \gamma (p_4,\lambda_4) \ \ , \label{gggg-process}
\eq are denoted as\footnote{Their sign is related to the sign of
the $S$-matrix through   $S_{\lambda_1 \lambda_2
\lambda_3\lambda_4}= 1+i (2\pi)^4 \delta(p_f-p_i)
F_{\lambda_1 \lambda_2 \lambda_3\lambda_4}$.} $F_{\lambda_1
\lambda_2 \lambda_3\lambda_4}(\s,\t,\u)$, where the momenta and
helicities of the incoming and out going photons are indicated in
parenthesis, and $\s=(p_1+p_2)^2$, $\t=(p_1-p_3)^2$,
$\u=(p_1-p_4)^2$.

Bose statistics demands
\bqa
F_{\lambda_1 \lambda_2 \lambda_3\lambda_4}(\s,\t,\u) &=&
F_{\lambda_2 \lambda_1 \lambda_4\lambda_3}(\s,\t,\u) \ ,
\label{Bose2} \\
F_{\lambda_1 \lambda_2 \lambda_3\lambda_4}(\s,\t,\u) &=&
F_{\lambda_2 \lambda_1 \lambda_3\lambda_4}(\s,\u,\t) \ ,
\label{Bose1}
\eqa
while  crossing symmetry implies
\bqa
F_{\lambda_1 \lambda_2 \lambda_3\lambda_4}(\s,\t,\u) =&
F_{-\lambda_4 \lambda_2 \lambda_3 -\lambda_1}(\t,\s,\u) =&
F_{\lambda_1 -\lambda_3 -\lambda_2\lambda_4}(\t,\s,\u) \ ,
\label{cross-ts}\\
F_{\lambda_1 \lambda_2 \lambda_3\lambda_4}(\s,\t,\u) =&
F_{-\lambda_3 \lambda_2 -\lambda_1\lambda_4}(\u,\t,\s) =&
F_{\lambda_1 -\lambda_4 \lambda_3 -\lambda_2}(\u,\t,\s) \ .
\label{cross-us}
\eqa
If parity and time inversion invariance holds, we have respectively
the additional constraints
\bqa
F_{\lambda_1 \lambda_2 \lambda_3\lambda_4}(\s,\t,\u) &=&
F_{-\lambda_1-\lambda_2- \lambda_3-\lambda_4}(\s,\t,\u) \ \ ,
 \label{parity} \\
F_{\lambda_3 \lambda_4 \lambda_1\lambda_2}(\s,\t,\u) & = &
F_{\lambda_1 \lambda_2 \lambda_3\lambda_4}(\s,\t,\u) \ \  .
\label{time-inv}
\eqa

As a result, the 16 possible helicity amplitudes may be expressed in terms
of just the three amplitudes
$F_{+++-}(\s,\t,\u)$, $F_{++--}(\s,\t,\u)$ and
$F_{++++}(\s,\t,\u)$ through \cite{Jikia}
\bqa
 F_{\pm\pm\mp\pm}(\s,\t,\u)&= &F_{\pm\mp\pm\pm}(\s,\t,\u)=
F_{\pm\mp\mp\mp}(\s,\t,\u)= F_{---+}(\s,\t,\u) \nonumber \\
& = &F_{+++-}(\s,\t,\u) \ , \label{+++-} \\
F_{--++}(\s,\t,\u)& = &F_{++--}(\s,\t,\u) \ , \label{++--} \\
 F_{\pm\mp\pm\mp}(\s,\t, \u)&=&F_{----}(\u,\t,\s)=
F_{++++}(\u,\t,\s) \ , \label{++++1} \\
 F_{\pm\mp\mp\pm}(\s,\t, \u)&=&F_{\pm\mp\pm\mp}(\s,\u, \t)=
F_{++++}(\t,\s, \u)= F_{++++}(\t,\u, \s) \ .
\label{++++2}
\eqa

Using the notation of \cite{Hagiwara} for the $B_0$, $C_0$ and $D_0$
1-loop functions first introduced by Passarino and Veltman
\cite{Passarino}, as well as the shorthand notation
\bqa
B_0(s)& \equiv  & B_0(12)=B_0(s;m,m) \ , \label{B0} \\
C_0(s) & \equiv & C_0(123)=C_0(0,0,s;m,m,m) \ , \label{C0} \\
D_0(s,t) & \equiv & D_0(1234)=D_0(0,0,0,0,s,t;m,m,m,m)=D_0(t,s)
\  \label{D0}
\eqa
suggested by the masslessness of the photons,
the $W$ loop contribution may be written as\footnote{The easiest way
to calculate this, is by using a non-linear gauge as
in \cite{Dicus}, in which the couplings $\gamma W^\pm\phi^\mp$,
$Z W^\pm\phi^\mp$ vanish. As a result, in each loop,
we always have propagators of the same mass.} \cite{Jikia}
\bqa
&& \frac{F^W_{++++} (\s,\t,\u)}{\alpha^2}= 12
-12 \left (1 + \frac{2\u}{\s}\right ) B_0(\u)
-12 \left (1 + \frac{2\t}{\s}\right ) B_0(\t) + \nonumber \\
&& \frac{24\mwd \t\u}{\s}D_0(\u,\t) +16 \left (1- \frac{3\mwd}{2\s}
-\frac{3\t\u}{4\s^2} \right ) [2\t C_0(\t) +2\u C_0(\u)-\t\u D_0(\t,\u)]
\nonumber \\
&& +8(\s-\mwd)(\s-3\mwd)[D_0(\s,\t)+ D_0(\s,\u)+D_0(\t,\u)] \ ,
\label{W++++} \\[0.5cm]
&& \frac{F^W_{+++-} (\s,\t,\u)}{\alpha^2}= -12
+24 \mw^4 [D_0(\s,\t)+ D_0(\s,\u)+D_0(\t,\u)]  \nonumber \\
&& + 12\mwd \s\t\u \left  [\frac{D_0(\s,\t)}{\u^2}+
\frac{D_0(\s,\u)}{\t^2}+\frac{D_0(\t,\u)}{\s^2} \right ]
\nonumber \\
&&-24\mwd \left (\frac{1}{\s} + \frac{1}{\t} +\frac{1}{\u} \right )
[\t C_0(\t) +\u C_0(\u) + \s C_0(\s)] \ , \label{W+++-} \\[0.5cm]
&& \frac{F^W_{++--} (\s,\t,\u)}{\alpha^2}=  -12
+24 \mw^4 [D_0(\s,\t)+ D_0(\s,\u)+D_0(\t,\u)] \ .
\label{W++--}
\eqa

Correspondingly, the contribution from the circulation in a
loop of a fermion of charge $Q_f$ and mass $m_f$ is \cite{vanderBij}
\bqa
&& \frac{F^f_{++++} (\s,\t,\u)}{\alpha^2 Q_f^4}=- 8
+8 \left (1 + \frac{2\u}{\s}\right ) B_0(\u)
+8 \left (1 + \frac{2\t}{\s}\right ) B_0(\t)  \nonumber \\
&& - 8\left ( \frac{\t^2+\u^2}{\s^2}-~ \frac{4m_f^2}{\s} \right )
[\t C_0(\t) +\u C_0(\u)]
+8 m_f^2 (\s -2 m_f^2 )[D_0(\s,\t) + D_0(\s,\u) ]
\nonumber \\
&& -4 \left [ 4m_f^4 -(2\s m_f^2 +\t\u)~ \frac{\t^2+\u^2}{\s^2}
+ \frac{4 m_f^2\t\u}{\s} \right ] D_0(\t,\u) \  , \label{f++++}
\eqa
\bqa
F^f_{+++-}(\s,\t,\u) &=& - ~\frac{2}{3} Q_f^4 \left \{
F^W_{+++-}(\s,\t,\u)~;~ \mw \to m_f \right \}\ , \label{f+++-}\\
F^f_{++--}(\s,\t,\u) &=& -~ \frac{2}{3} Q_f^4  \left \{
F^W_{++--}(\s,\t,\u)~;~ \mw \to m_f \right \} \ . \label{f++--}
\eqa

Equations (\ref{W++++}-\ref{f++--}) are sufficient for calculating any
amplitude for the process (\ref{gggg-process}) in SM. For the  SUSY case
though, we also need the contributions to $F_{+++-}(\s,\t,\u)$,
$F_{++--}(\s,\t,\u)$ and $F_{++++}(\s,\t,\u)$ from a charged scalar
particle (\eg\@ a squark or slepton),
circulating in the loop. Thus, for a scalar particle
with charge $Q_{\tilde l}$ and mass $\msl$ we find
\bqa
&& \frac{F^{\slepton}_{++++} (\s,\t,\u)}{\alpha^2 Q_{\slepton}^4}= 4
-4 \left (1 + \frac{2\u}{\s}\right ) B_0(\u)
-4 \left (1 + \frac{2\t}{\s}\right ) B_0(\t) + \nonumber \\
&& \frac{8\msld \t\u}{\s}D_0(\t,\u) -~\frac{8\msld}{\s} \left (1+~
\frac{\u\t}{2\msld\s}  \right ) [2\t C_0(\t) +2\u C_0(\u)-\t\u D_0(\t,\u)]
\nonumber \\
&& +8\msl^4[D_0(\s,\t)+ D_0(\s,\u)+D_0(\t,\u)] \ , \label{slepton++++}
\eqa
\bqa
F^{\slepton}_{+++-}(\s,\t,\u) &=& \frac{1}{3} Q_{\slepton}^4 \left \{
F^W_{+++-}(\s,\t,\u)~;~ \mw \to \msl \right \}\ , \label{slepton+++-}\\
F^{\slepton}_{++--}(\s,\t,\u) &=& \frac{1}{3} Q_{\slepton}^4  \left \{
F^W_{++--}(\s,\t,\u)~;~ \mw \to \msl \right \} \ . \label{slepton++--}
\eqa

\renewcommand{\theequation}{B.\arabic{equation}}
\renewcommand{\thesection}{B.\arabic{section}}
\setcounter{equation}{0}
\setcounter{section}{0}

\vspace*{3cm}

{\large \bf Appendix B: Density matrix of a pair of
 backscattered photons.}

Following \cite{LCgg}, we collect in this appendix
the formulae describing the helicity density
matrix of the photon pair produced by backscattering of two
laser photons from the corresponding highly energetic
$e^\pm$ beams of the Linear Collider.\par

We denote by $E$ the energy of each  incoming $e^\pm$ beam,
while $P_e=2\lambda_e$
describes its longitudinal polarization, and $\lambda_e$
is its average helicity. An $e^\pm$ beam is assumed to
collide with a laser
photon moving along the opposite direction with energy
$\omega_0$. In its helicity basis, each laser photon is characterized
by a normalized density matrix of the form
\bq
\label{laser-rho}
\rho_{laser}^N ~ = ~\frac{1}{2}
\left (\matrix{1+P_\gamma & -P_t e^{-2i\phi} \cr
-P_t e^{+2i\phi} & 1-P_\gamma } \right ) \ \ .
\eq
$P_\gamma$ describes the average helicity of the laser photon,
while $P_t$ ($P_t \geq 0 $) denotes its  maximum average transverse
 polarization along a direction determined by the azimuthal angle
$\phi$. This $\phi$ angle is defined with respect to a $\hat z$-axis
pointing {\it opposite} to the laser momentum; \ie~ along the
direction that the backskattered photon moves. By definition
\bq
\label{PtPgineq}
0 \leq P_\gamma^2+P_t^2 \leq 1 \ \ .
\eq\par

After the Compton scattering of $e^\pm$ from the laser photon,
the electron beam looses most of its energy and a beam of "backscattered
photons" is
produced, moving essentially along the direction of the original
$e^\pm$ momentum and characterized, in its helicity basis,
by the density matrix
\bqa
\rho^B &=& \frac{dN}{dx}~\rho^{BN} \ \ , \label{rhoB} \\
\rho^{BN} & =& \frac{1}{2}
\left (\matrix{1+\xi_2(x) & - \xi_{3}(x) e^{-2i\phi} \cr
- \xi_{3}(x)  e^{+2i\phi} & 1-\xi_2(x) } \right ) \ \ ,
\label{rhoBN}
\eqa
where $x \equiv \omega/E$ and $x_0 \equiv 4E \omega_0/m_e^2$; with
$\omega$ being the energy of the back-scattered photon,
and $\omega_0$ and $E$ as  defined above.
These satisfy the kinematical constraints
\bq
\label{laser-kin}
0 \leq x \leq x_{max} ~\equiv~ \frac{x_0}{1+x_0} \ \ \ \ ,
\ \ \ \ 0\leq x_0 \leq 2 (1+\sqrt 2) \ \ .
\eq
We also note from (\ref{rhoBN}, \ref{laser-rho}),  that the
azimuthal angles of the
maximum average transverse polarizations of the backscattered
and laser photons are the same, when defined around the momentum of the
backscattered photon  \cite{LCgg}. Moreover,
in analogy to (\ref{PtPgineq}), we also have
\bq
0 \leq \xi_2^2(x)+ \xi_{3}^2(x) \leq 1 \ \ , \ \ (\xi_3 \geq 0) \ \ .
\eq\par

In (\ref{rhoB}),  $\rho^{BN}$ is the normalized
 density matrix of a backscattered photon,  ($Tr\rho^{BN}=1$);
while $dN/dx$ is the overall flux of backscattered photons,
per unit of $x$ and  unit  $e^\pm$ flux.
Their form, immediately after the
production of the backscattered photon  at the
{\it conversion point}, is given by \cite{LCgg, gamma97}
\bqa
\frac{dN(x)}{dx} &= & \frac{\C(x)}{\D(x_0)}\ \ , \label{gamma-flux} \\
\C(x) &=& f_0(x)+P_eP_\gamma f_1(x) \ \ , \label{laser-C} \\
\D(x_0) & = & \D_0 (x_0) + P_eP_\gamma \D_1 (x_0) \ \ ,
\label{laser-D} \\
\xi_2(x) & = & \frac{P_e f_2(x) +P_\gamma f_3(x)}{C(x)} \ \ ,
\label{laser-xi2}\\
\xi_{3}(x) & =& \frac{2 r^2(x) P_t}{\C(x)} \ \ , \label{laser-xi13}
\eqa
where  $f_i(x)$, $\D_j(x_0)$ are given in \cite{LCgg, Tsi}.

If both $e^-e^+$ beams of the Linear Collider are
transformed to photons,  by applying
two lasers working respectively with   parameters
$x_0$ and $x_0^\prime$ (compare \ref{laser-kin});
then the (unnormalized) density matrix of the photon pair
in their helicity basis
$\R_{\mu_1\mu_2 ; \tilde \mu_1 \tilde \mu_2}$, is determined by
$\rho^B$, $\rho^{\prime B}$ via (compare (\ref{rhoB}))
\bqa
\label{R-matrix}
\frac{d}{d\tau}\,
\R_{\mu_1\mu_2 ; \tilde \mu_1 \tilde \mu_2}(\tau)
& = & \rho^B_{\mu_1 \tilde \mu_1} \bigotimes
{\rho}^{\prime B}_{\mu_2 \tilde \mu_2} \equiv
\int_{\frac{\tau}{x_{max}^\prime}}^{x_{max}}
\frac{dx}{x} \rho^B_{\mu_1 \tilde \mu_1}(x)
{\rho}^{\prime B}_{\mu_2 \tilde \mu_2} \left (\frac{\tau}{x}
\right )\ , \nonumber \\
&\equiv & \frac{d\bar L_{\gamma \gamma}(\tau)}{d\tau}\ \langle
\rho^{BN}_{\mu_1 \tilde \mu_1}
{\rho}^{\prime BN}_{\mu_2 \tilde \mu_2} \rangle \ \ ,
\eqa
where
\bq
\tau ~ \equiv ~ \frac{s_{\gamma \gamma}}{s_{ee}} \ \ ,
\eq
with $s_{ee}$ and $s_{\gamma \gamma}$ being the squares of the
c.m. energies of the $e^-e^+$ and $\gamma \gamma $ systems
respectively. In the r.h.s. of (\ref{R-matrix}),
$d\bar L_{\gamma \gamma}/d\tau$ is the overall $\gamma
\gamma $ luminosity per unit $e^-e^+$ flux,  defined
by the convolution of the two  $\gamma $ luminosities
given  in (\ref{gamma-flux}).
Thus, if the {\it conversion} points
where each of the two photons are produced through laser backscattering,
coincide with their
{\it interaction} point, then
\bq
\label{Lgammagamma}
\frac{d\bar L_{\gamma \gamma}}{d\tau} ~=~
\frac{1}{\D(x_0)\D^\prime(x_0^\prime)}
\int_{\frac{\tau}{x_{max}^\prime}}^{x_{max}}
\frac{dx}{x} \C(x) \C^\prime \left( \frac{\tau}{x} \right )
~\equiv ~ \frac{1}{\D(x_0)\D^\prime(x_0^\prime)}
(\C \bigotimes \C^\prime) \ ,
\eq
where $\C,~\D$ and $\C^\prime, ~\D^\prime$ are determined through
(\ref{laser-C}, \ref{laser-D}) by the polarization and the
 $x_0$ and  $x_0^\prime$ parameters of the two photons. The later
parameters
also determine $x_{max}$ and  $x_{max}^\prime$ respectively;
(compare (\ref{laser-kin})). Finally, the definition of the
{\it average} $\langle \rho^{BN}_{\mu_1 \tilde \mu_1}
{\rho}^{\prime BN}_{\mu_2 \tilde
\mu_2} \rangle$ appearing in the r.h.s. of
(\ref{R-matrix}) for the two photons, implies also
the definitions
\bqa
\langle \xi_i  \xi_j^\prime  \rangle & = &
\frac{(\C \xi_i \bigotimes \C^\prime \xi_j^\prime )}
{\C \bigotimes \C^\prime} \ \ ,
\label{xixjprime-ave}\\
\langle \xi_i \rangle  =
\frac{(\C \xi_i) \bigotimes \C^\prime }{\C \bigotimes \C^\prime}
& , &
\langle \xi_i^\prime \rangle  =
\frac{\C  \bigotimes ( \C^\prime \xi_i^\prime)}{\C \bigotimes \C^\prime}
\ \ , \label{xi-ave}
\eqa
where the same notation as in the r.h.s. of (\ref{Lgammagamma})
has been used. \par

The results for various polarizations of the $e^\pm$ beams and the
laser photons, and various values of the $x_0$, $x_0^\prime$
parameters,  are indicated in Figure \ref{flux-dis}a,b.

\clearpage
\newpage

\clearpage
\newpage

\addtocounter{figure}{-1}

\clearpage
\newpage

\begin{figure}[p]
\vspace*{-4cm}
\[
\epsfig{file=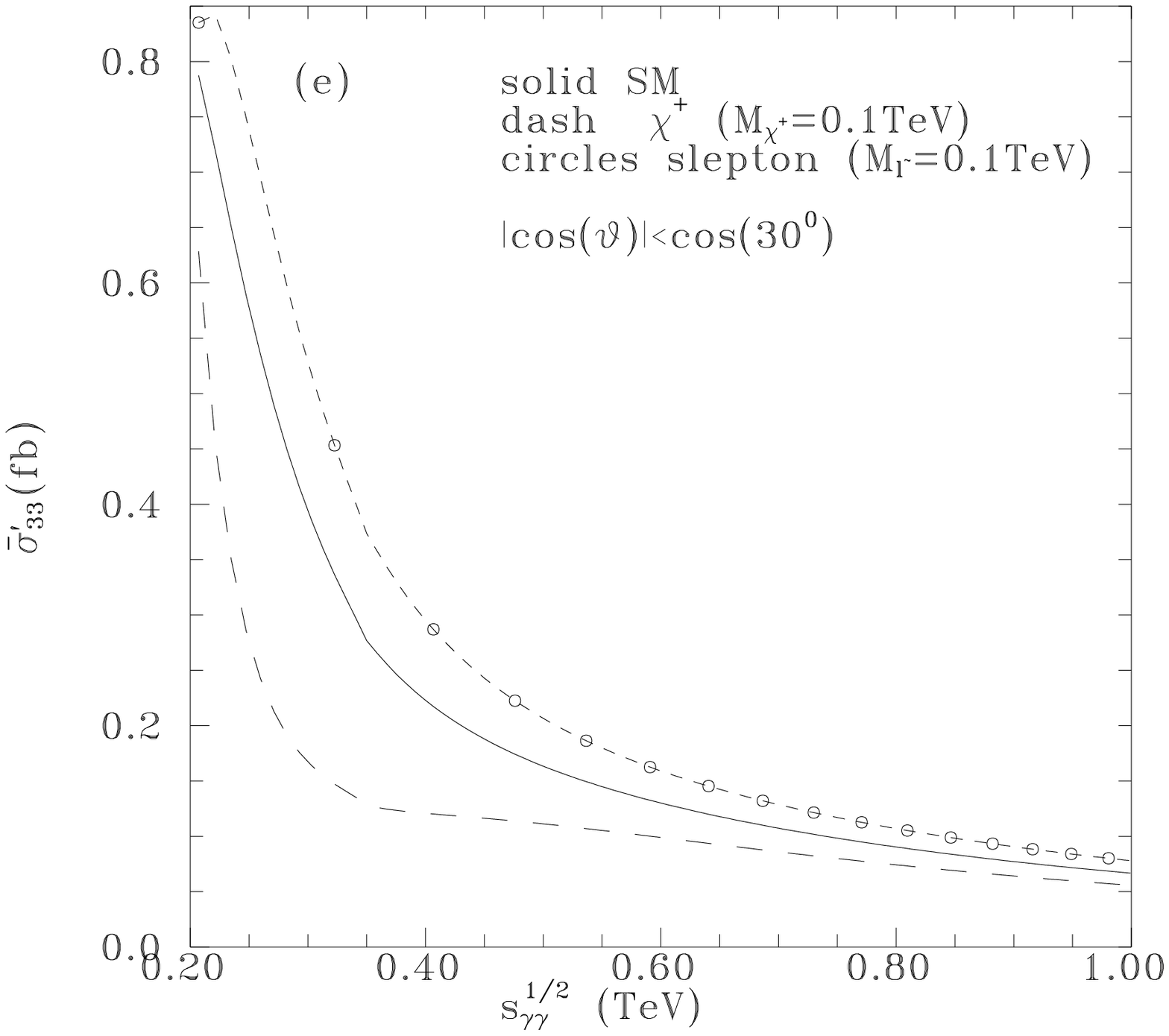,height=7.5cm}\hspace{0.5cm}
\epsfig{file=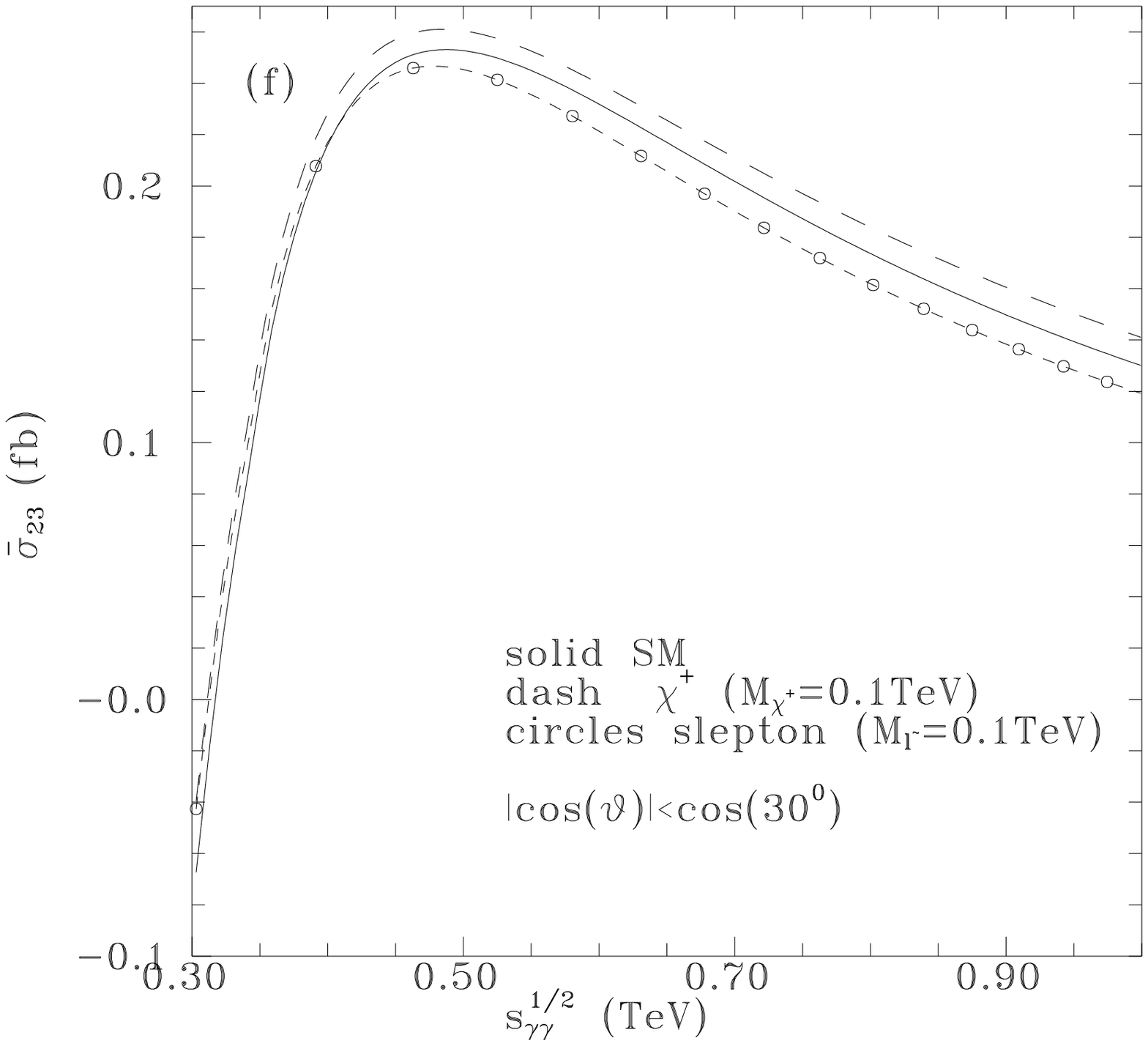,height=7.5cm}
\]
\vspace*{1.cm}
\caption[1]{ $\bar \sigma_{33}^\prime $ and
$\bar \sigma_{23}$ for SM (solid) and in the presence
of a chargino (dash) or a charged slepton (circles) contribution.}
\label{SUSY-sig33p-23}
\end{figure}

\clearpage
\newpage

\begin{figure}[p]
\vspace*{-3cm}
\[
\epsfig{file=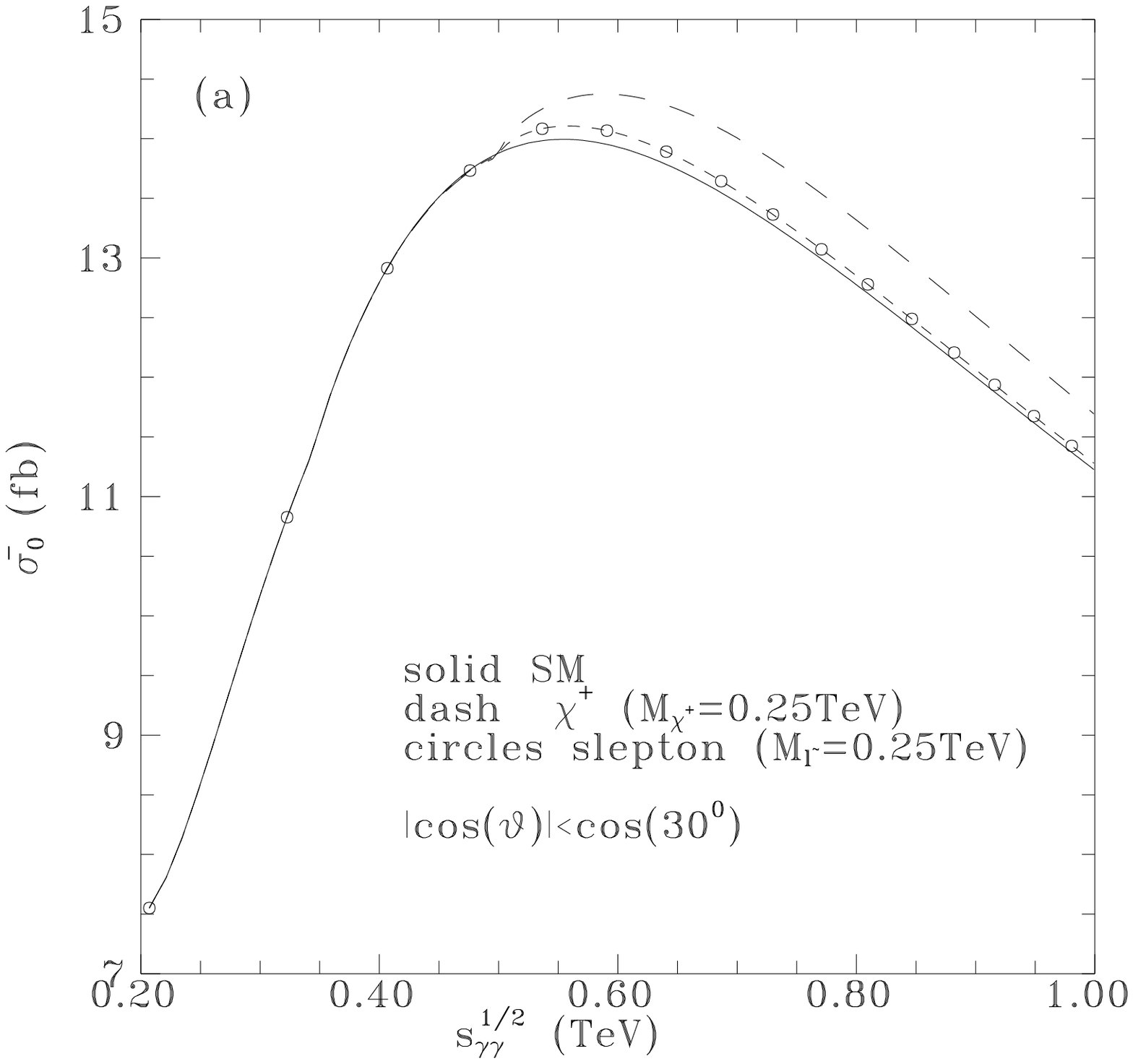,height=7.5cm}\hspace{0.5cm}
\epsfig{file=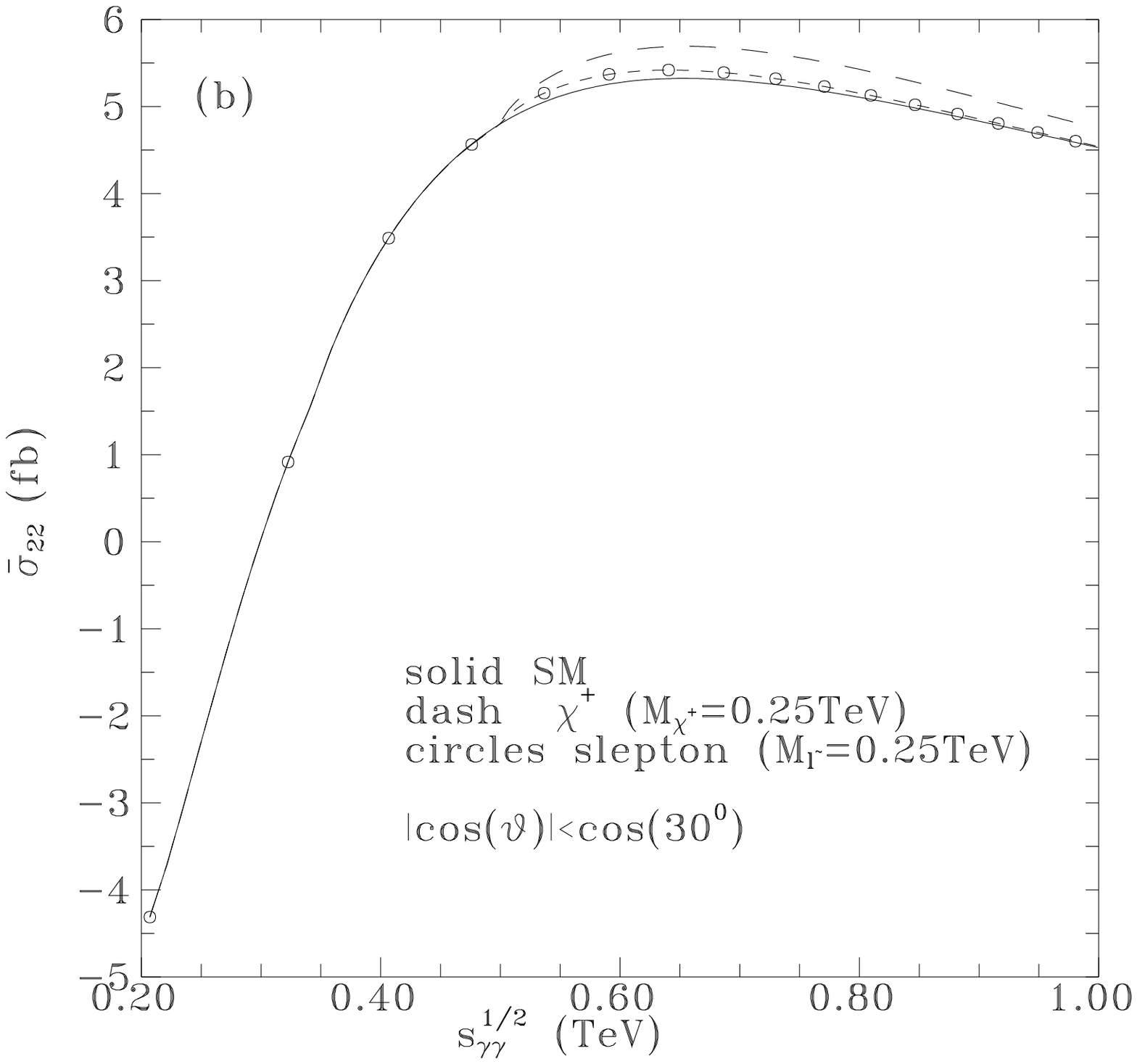,height=7.5cm}
\]
\vspace*{1.5cm}
\[
\epsfig{file=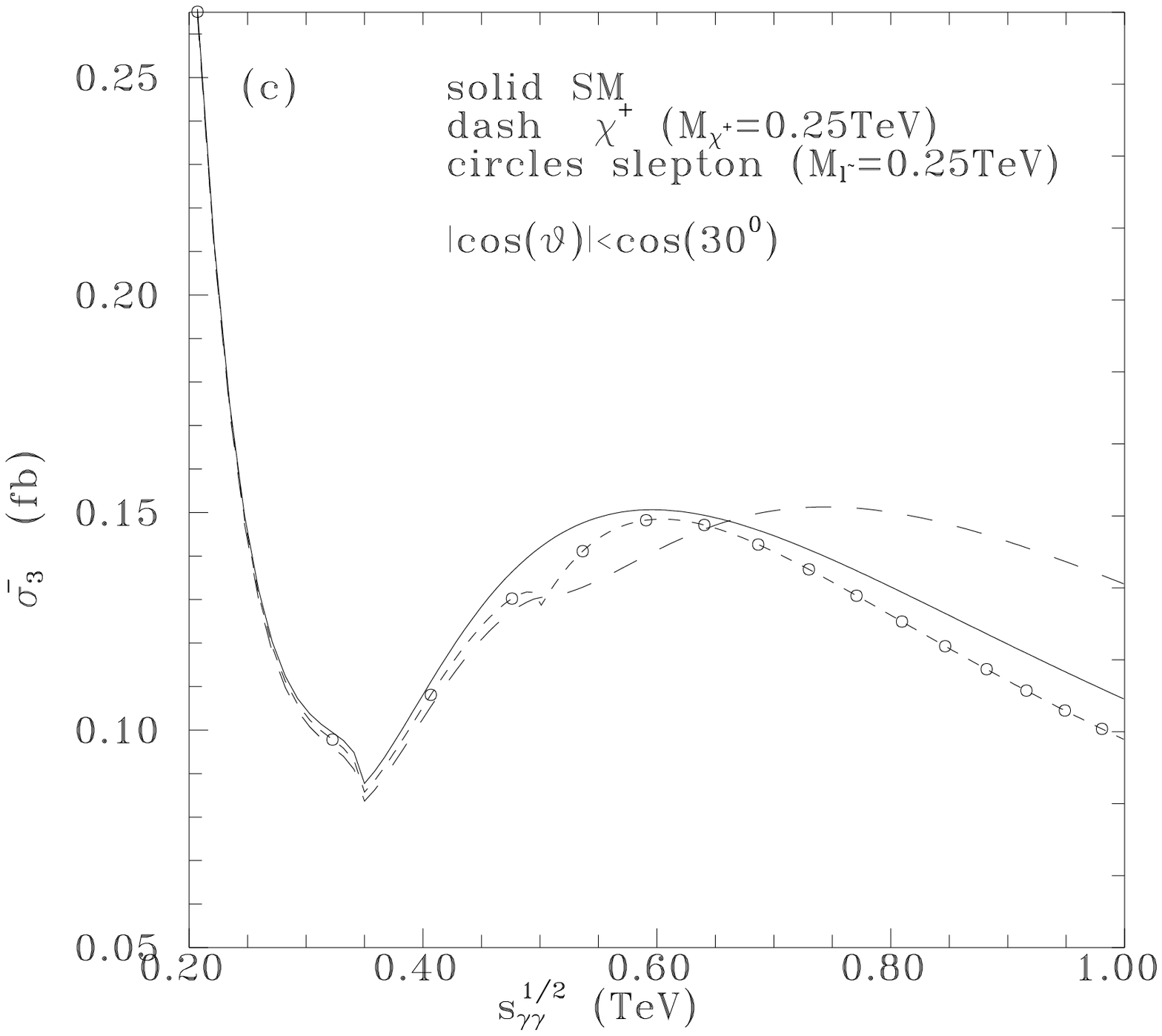,height=7.5cm}\hspace{0.5cm}
\epsfig{file=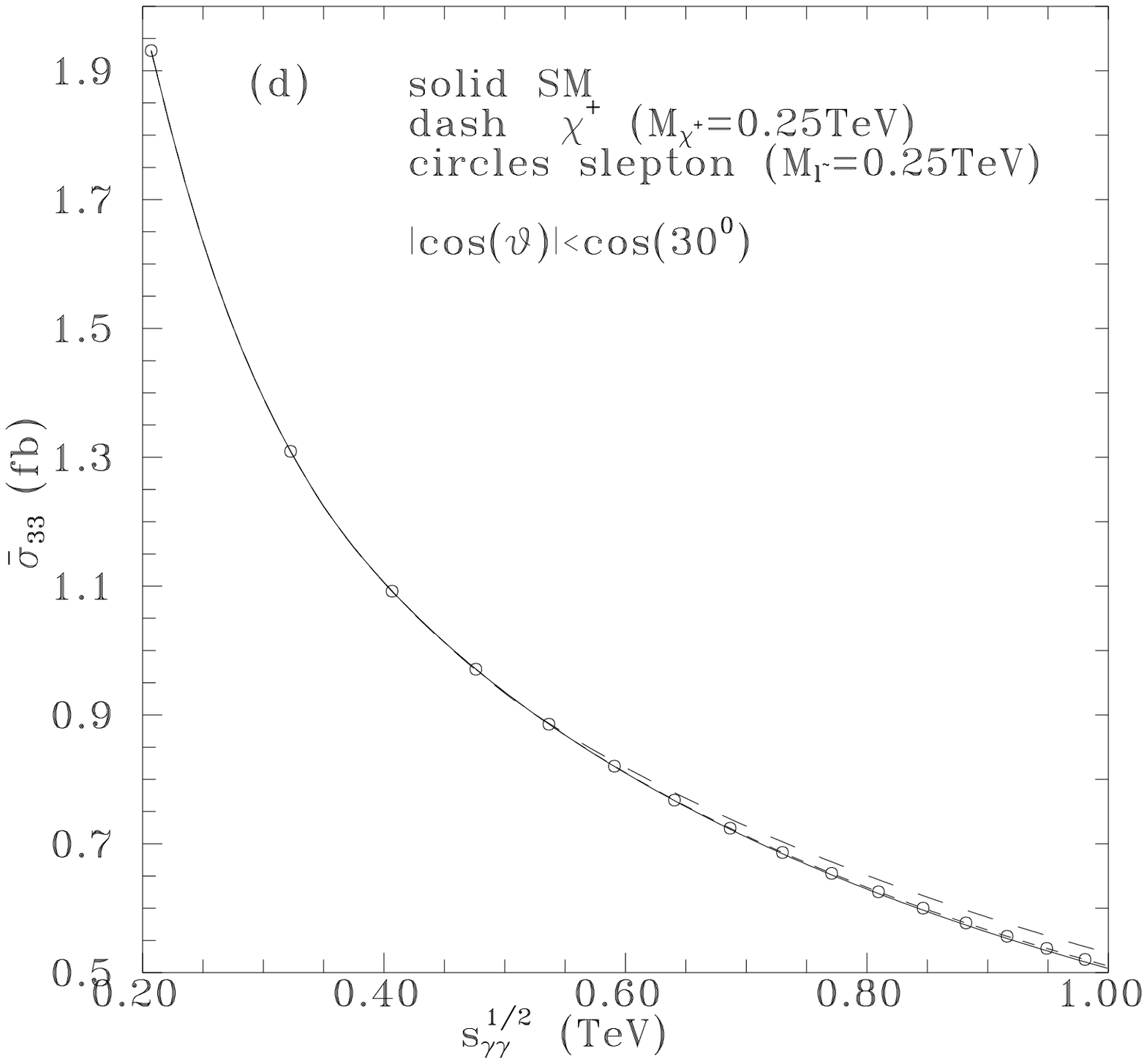,height=7.5cm}
\]
\vspace*{1.cm}
\caption[1]{$\bar \sigma_0$, $\bar \sigma_{22}$, $\bar \sigma_3 $
and  $\bar \sigma_{33} $ for SM (solid) and in the presence
of a chargino (dash) or a charged slepton (circles) contribution. }
\label{SUSY-sig0-33-2}
\end{figure}

\addtocounter{figure}{-1}

\clearpage
\newpage

\begin{figure}[p]
\vspace*{-4cm}
\[
\epsfig{file=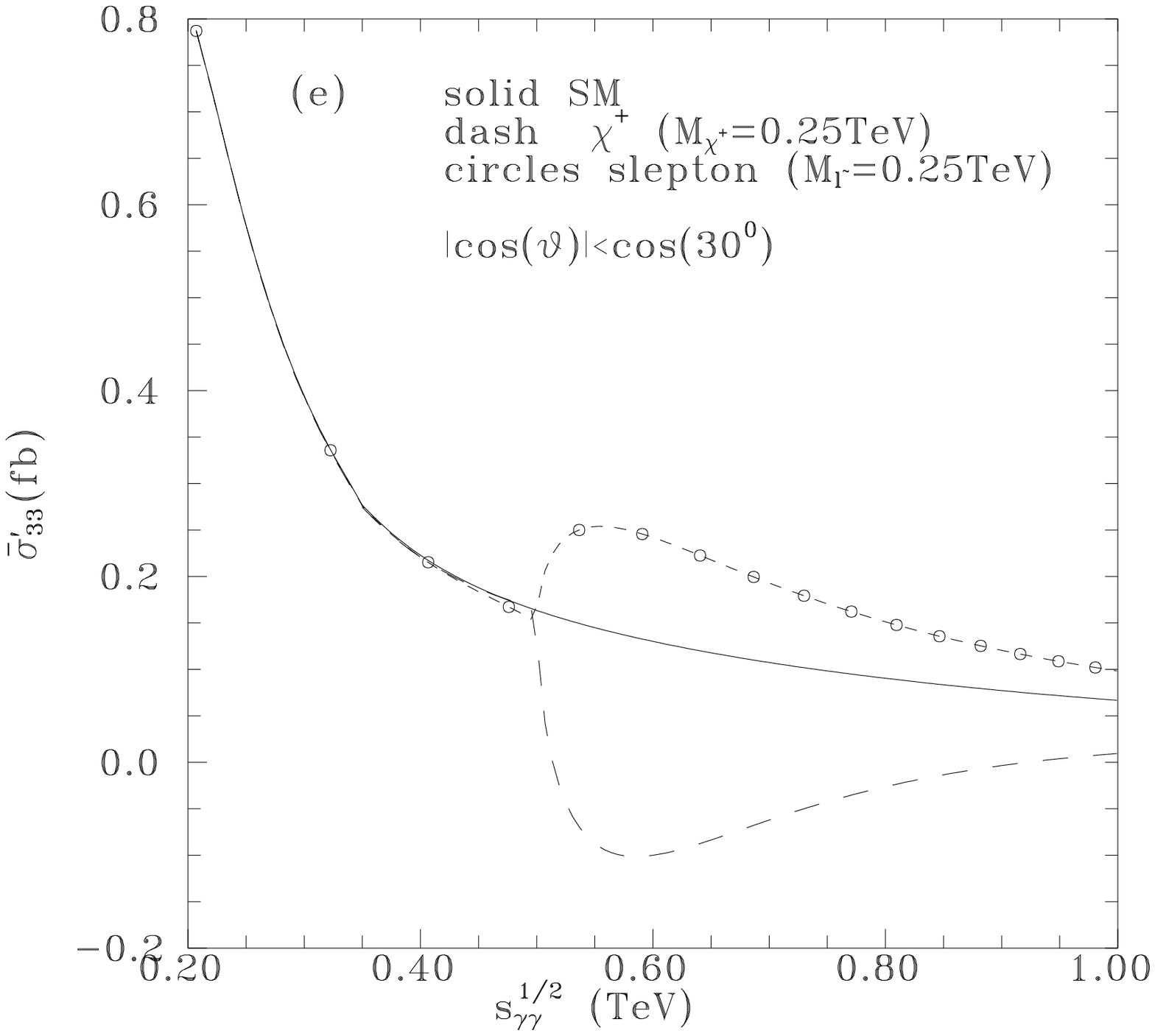,height=7.5cm}\hspace{0.5cm}
\epsfig{file=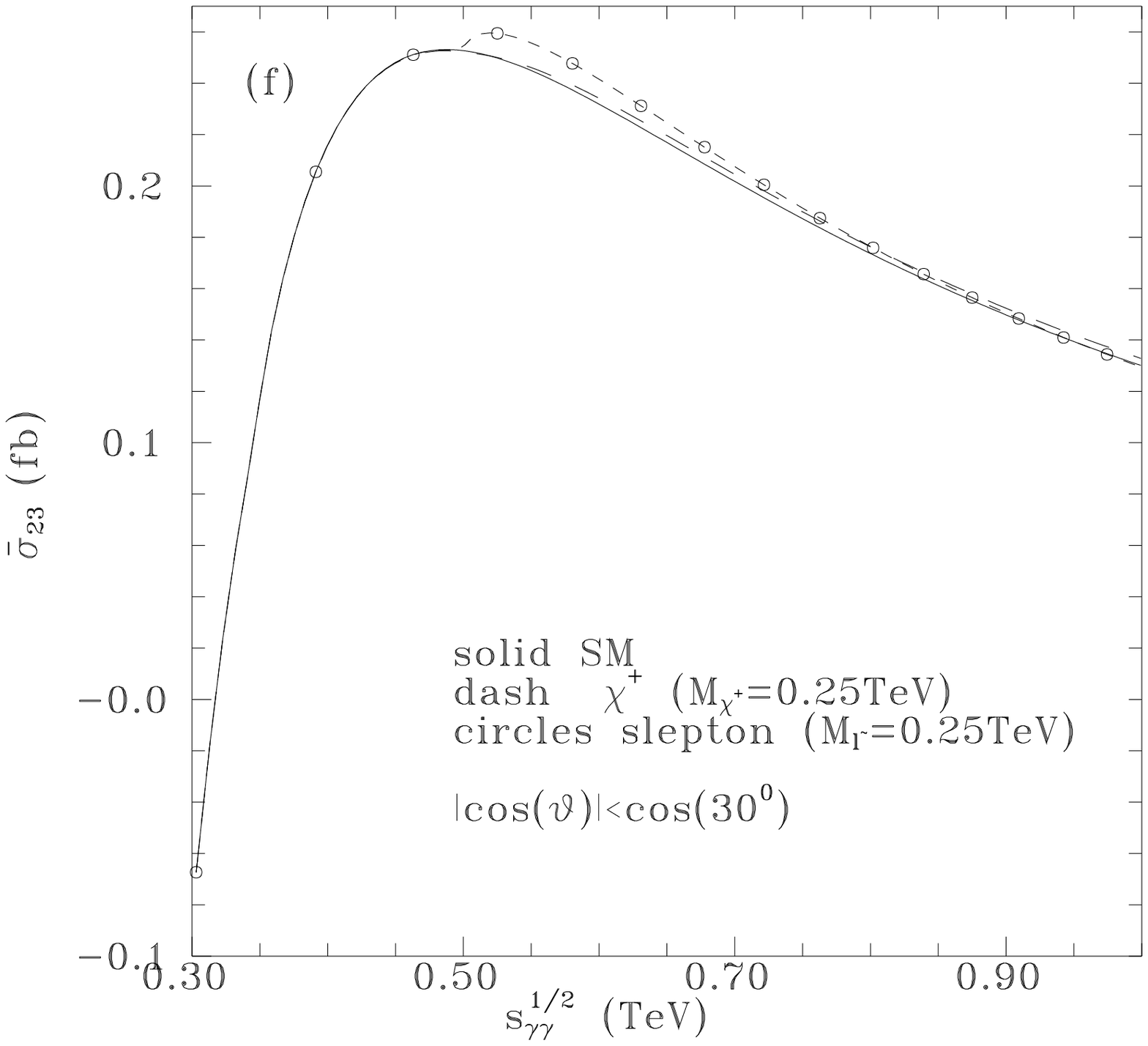,height=7.5cm}
\]
\vspace*{1.cm}
\caption[1]{ $\bar \sigma_{33}^\prime $ and
$\bar \sigma_{23}$ for SM (solid) and in the presence
of a chargino (dash) or a charged slepton (circles) contribution.}
\label{SUSY-sig33p-23-2}
\end{figure}

\clearpage
\newpage

\begin{figure}[p]
\vspace*{-4cm}
\[
\epsfig{file=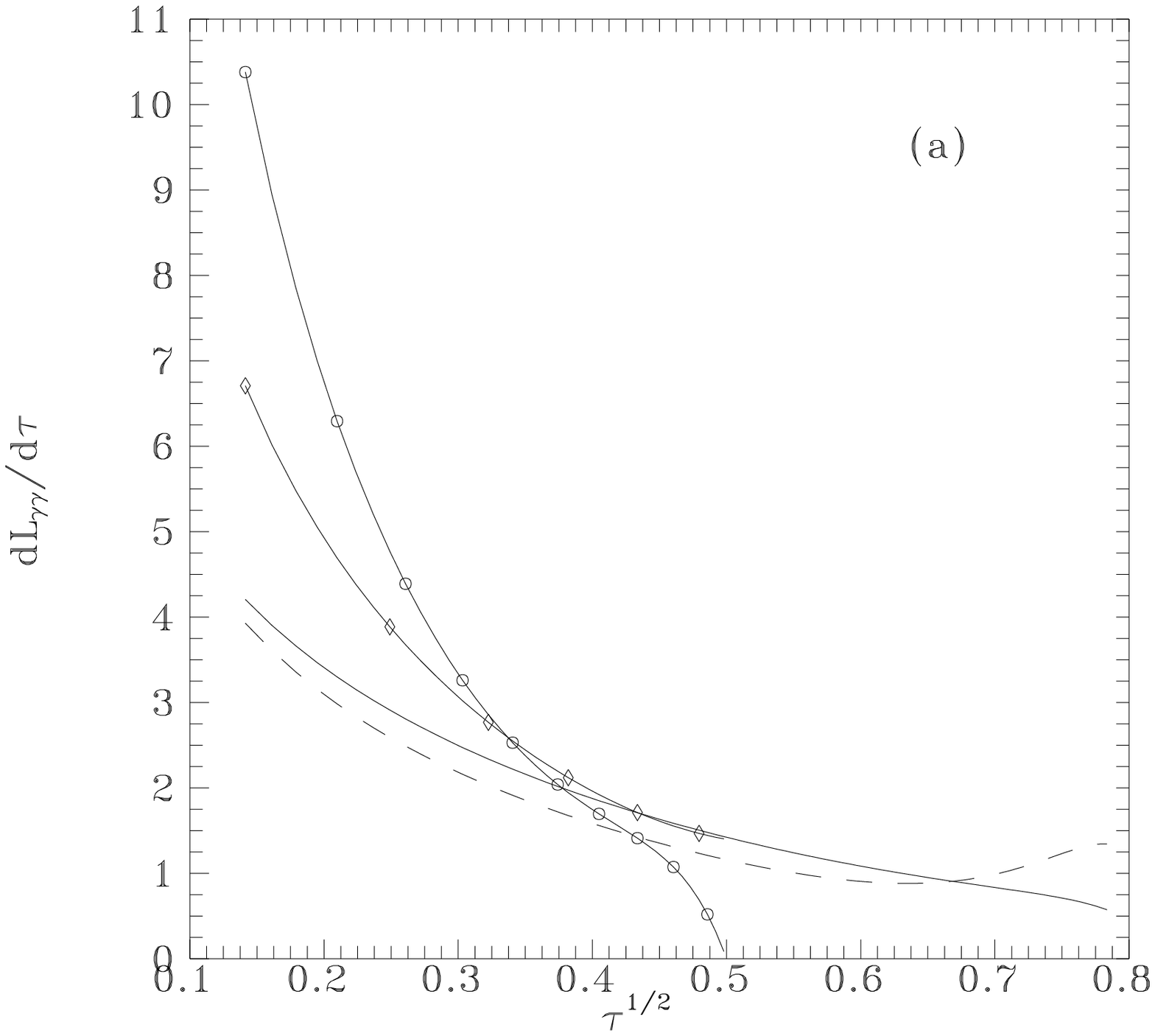,height=7.5cm}\hspace{0.5cm}
\epsfig{file=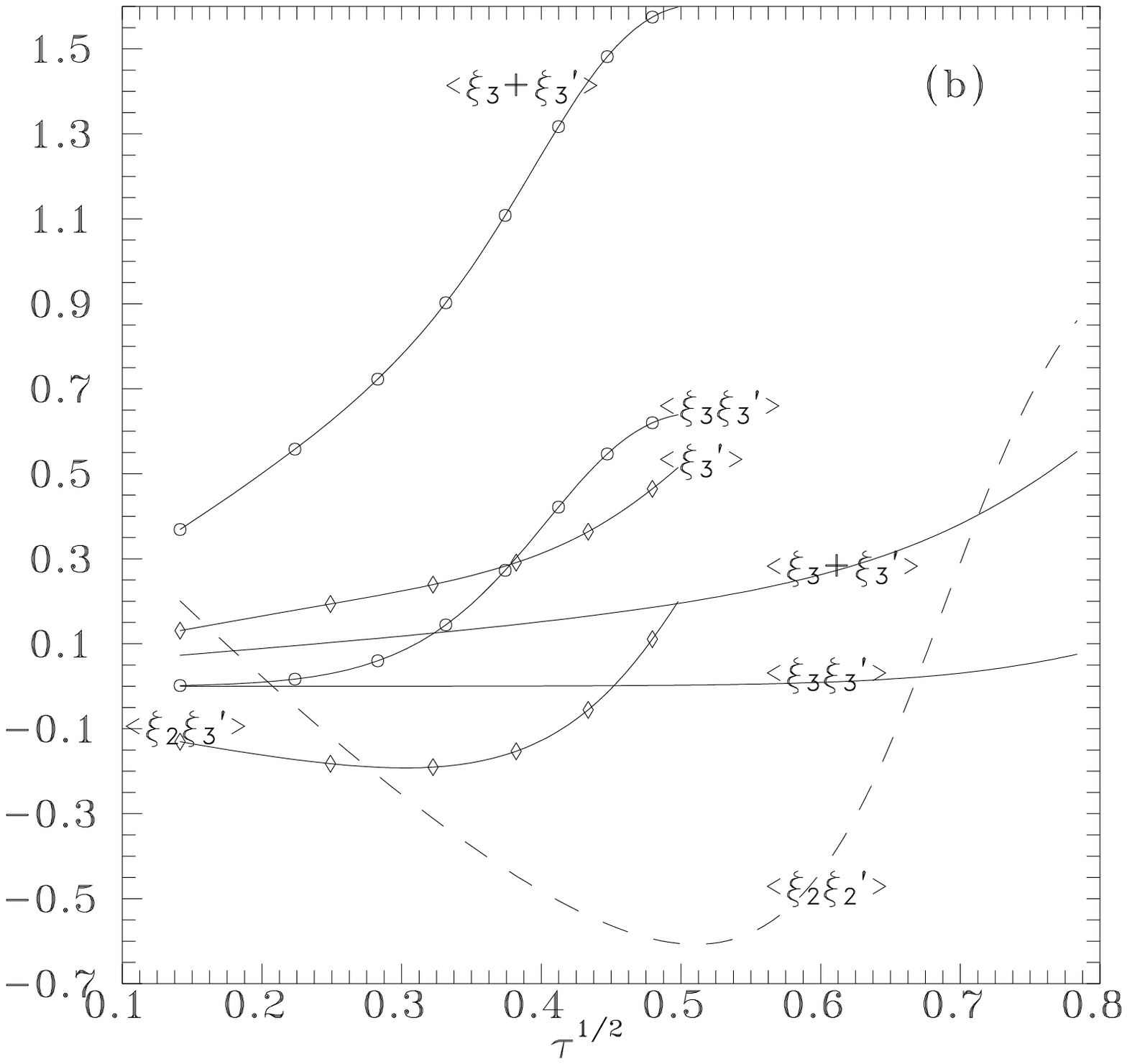,height=7.5cm}
\]
\vspace*{1.cm}
\caption[1]{Overall flux factor (a) and elements of the normalized
density matrix (b), for the two backscattered photons
with $P_e=P_e^\prime=0.8$, $P_\gamma
=P_\gamma^\prime=-1$, $P_t=P_t^\prime=0$, $x_0=x_0^\prime=4.83$, (dash);
$P_e=P_e^\prime= P_\gamma=P_\gamma^\prime=0$, $P_t=P_t^\prime=1$,
and $x_0=x_0^\prime=4.83$ (solid) or $x_0=x_0^\prime=1$ (circles);
$P_e=0.8$, $P_\gamma=-1$, $P_t=0$ $x_0=4.83$,
$P_e^\prime=0$, $P_\gamma^\prime=0$, $P_t^\prime=1$ $x_0^\prime=1$,
(rhombs).}
\label{flux-dis}
\end{figure}

\clearpage
\newpage

\begin{figure}[p]
\vspace*{-3cm}
\[
\epsfig{file=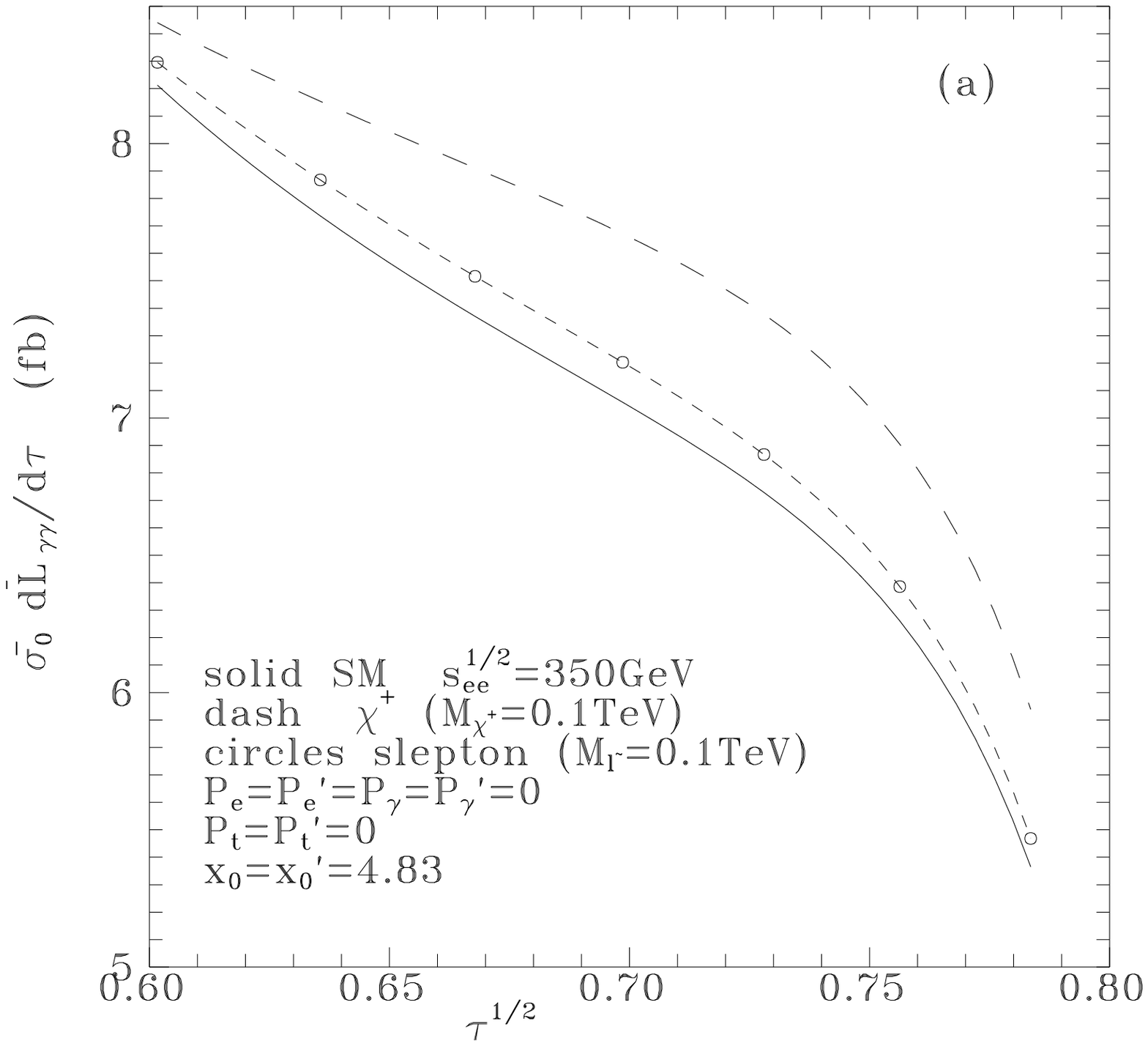,height=7.5cm}\hspace{0.5cm}
\epsfig{file=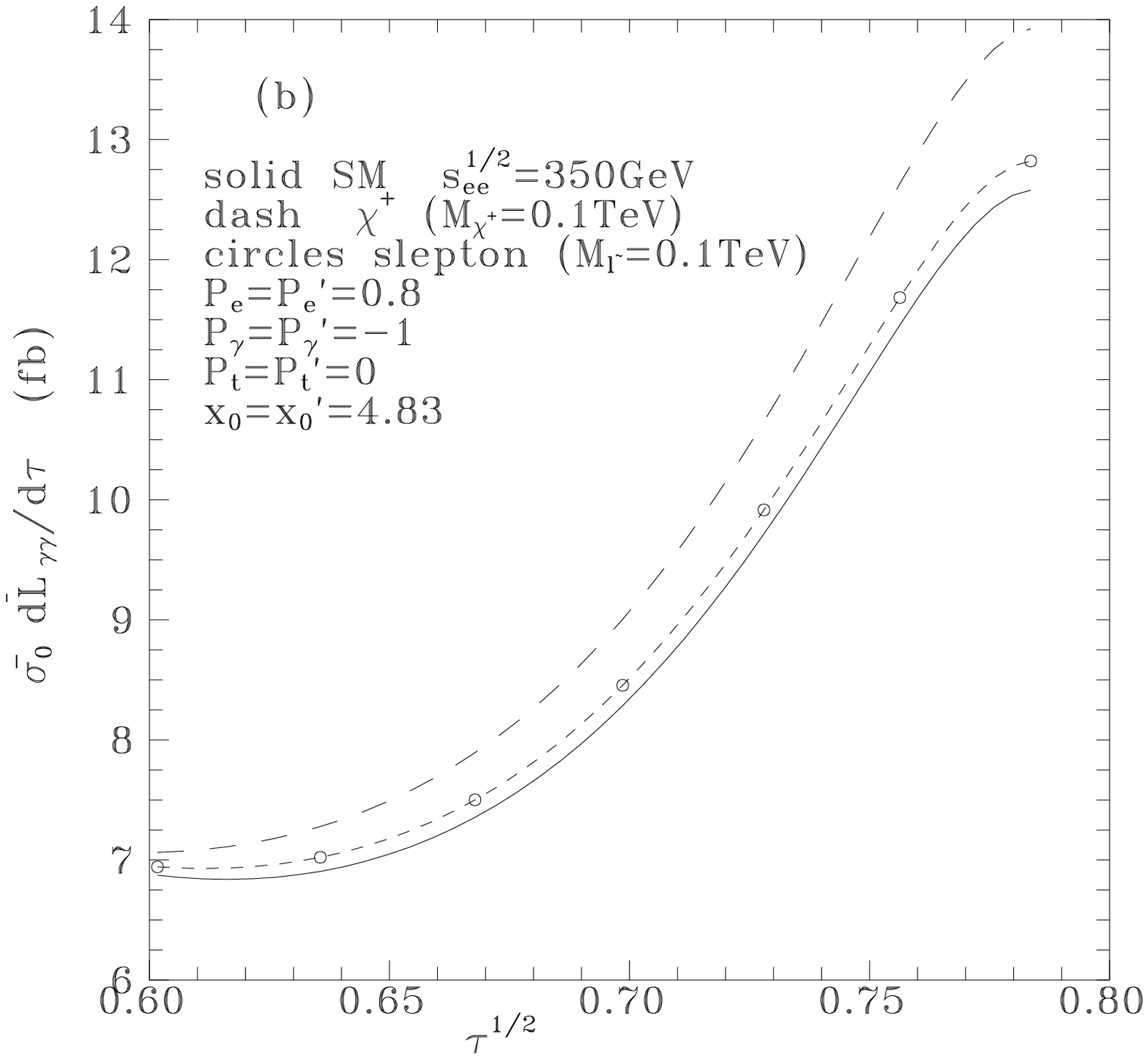,height=7.5cm}
\]
\vspace*{1.5cm}
\[
\epsfig{file=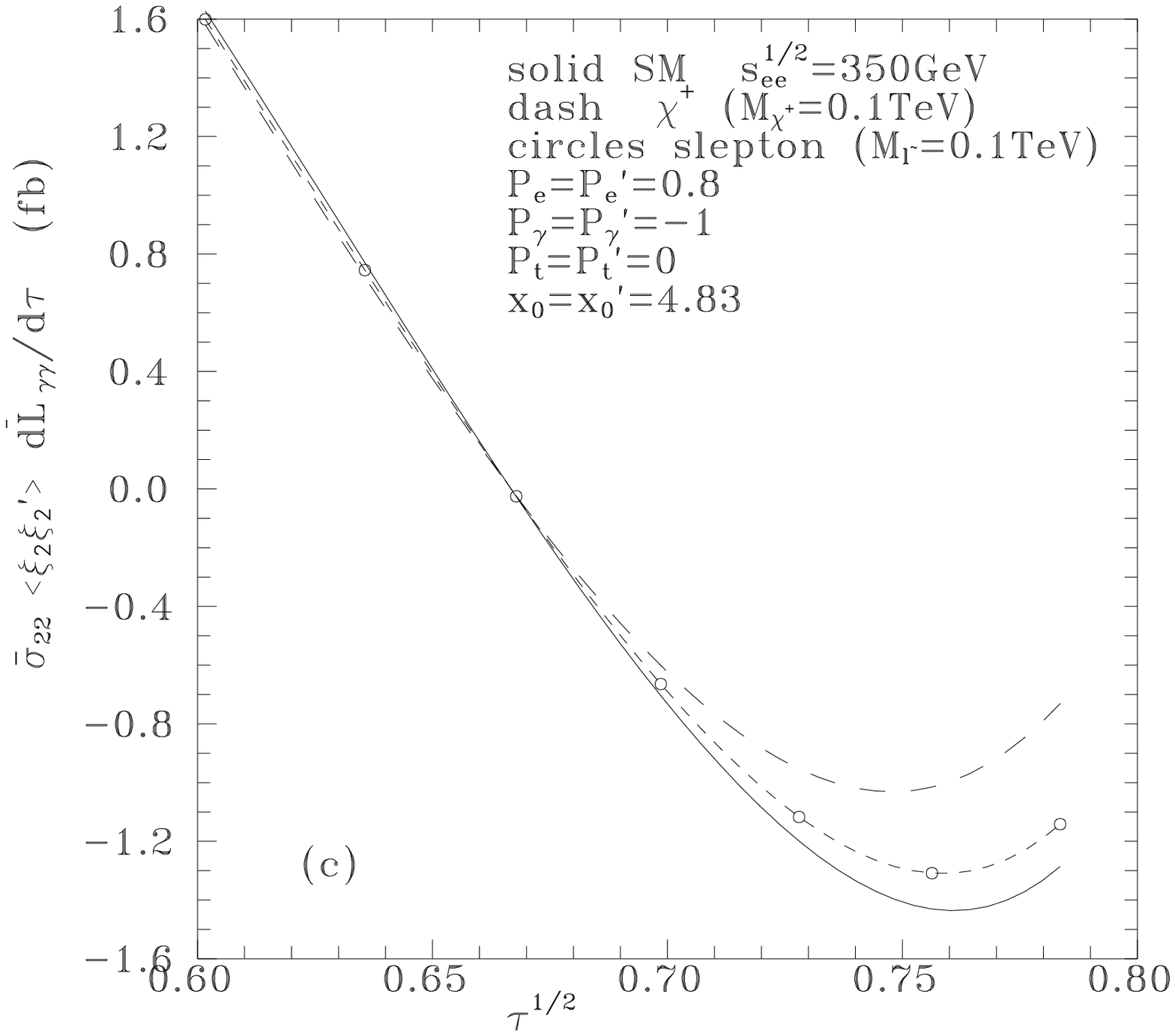,height=7.5cm}
\]
\vspace*{1.cm}
\caption[1]{$\bar \sigma_0$ and ($\bar \sigma_0$, $\bar
\sigma_{22}$) cross sections integrated over
$|\cos(\vartheta^*)|<\cos(30^0)$, multiplied by the indicated
photon density matrix elements for the indicated polarizations
 and $x_0$, $x_0^\prime$ parameters.
The SM and SUSY contributions induced by one chargino or
one charged slepton  with mass of 100 GeV, are also indicated.}
\label{SUSY-sig-flux1}
\end{figure}

\clearpage
\newpage

\begin{figure}[p]
\vspace*{-3cm}
\[
\epsfig{file=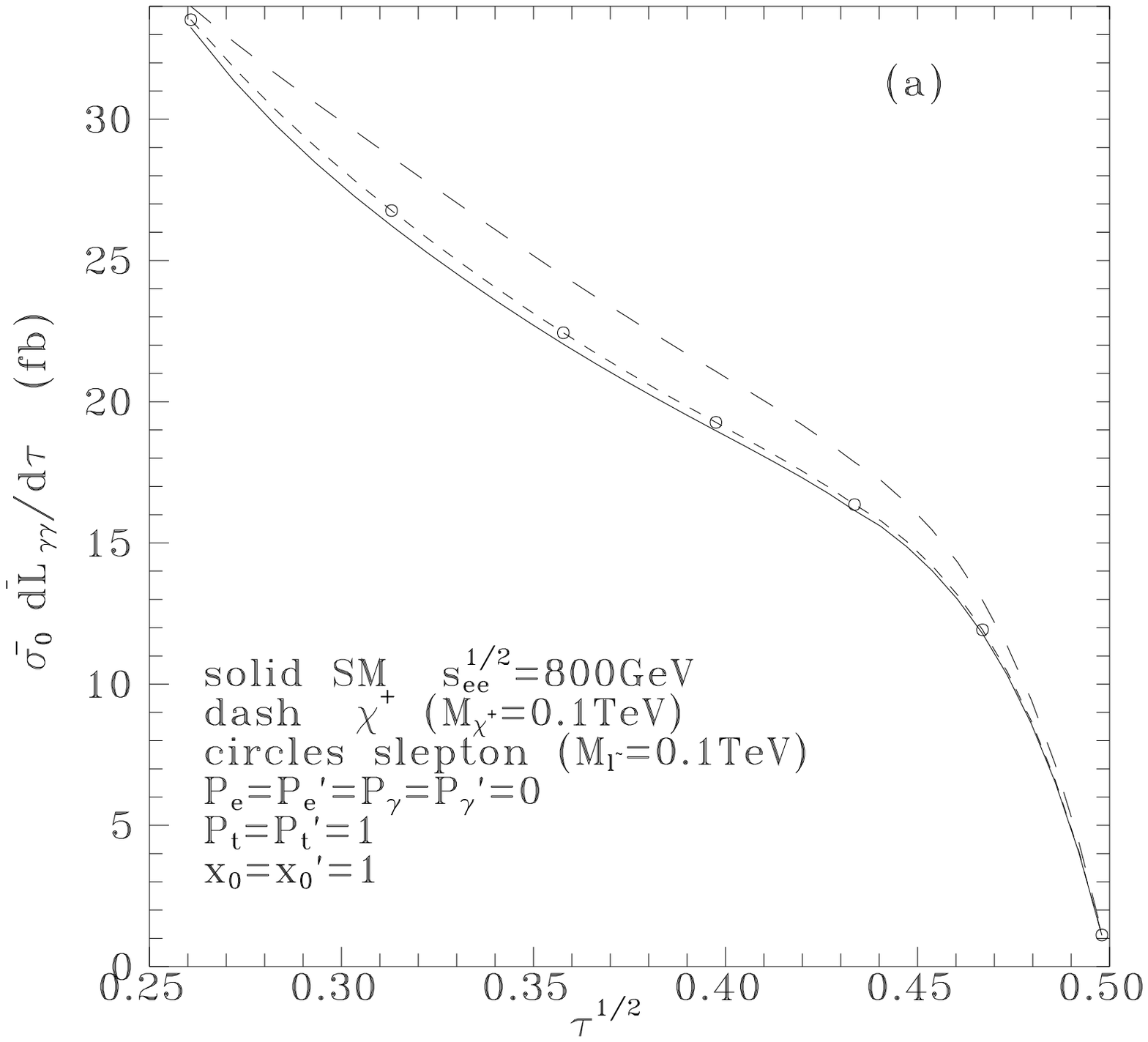,height=7.5cm}\hspace{0.5cm}
\epsfig{file=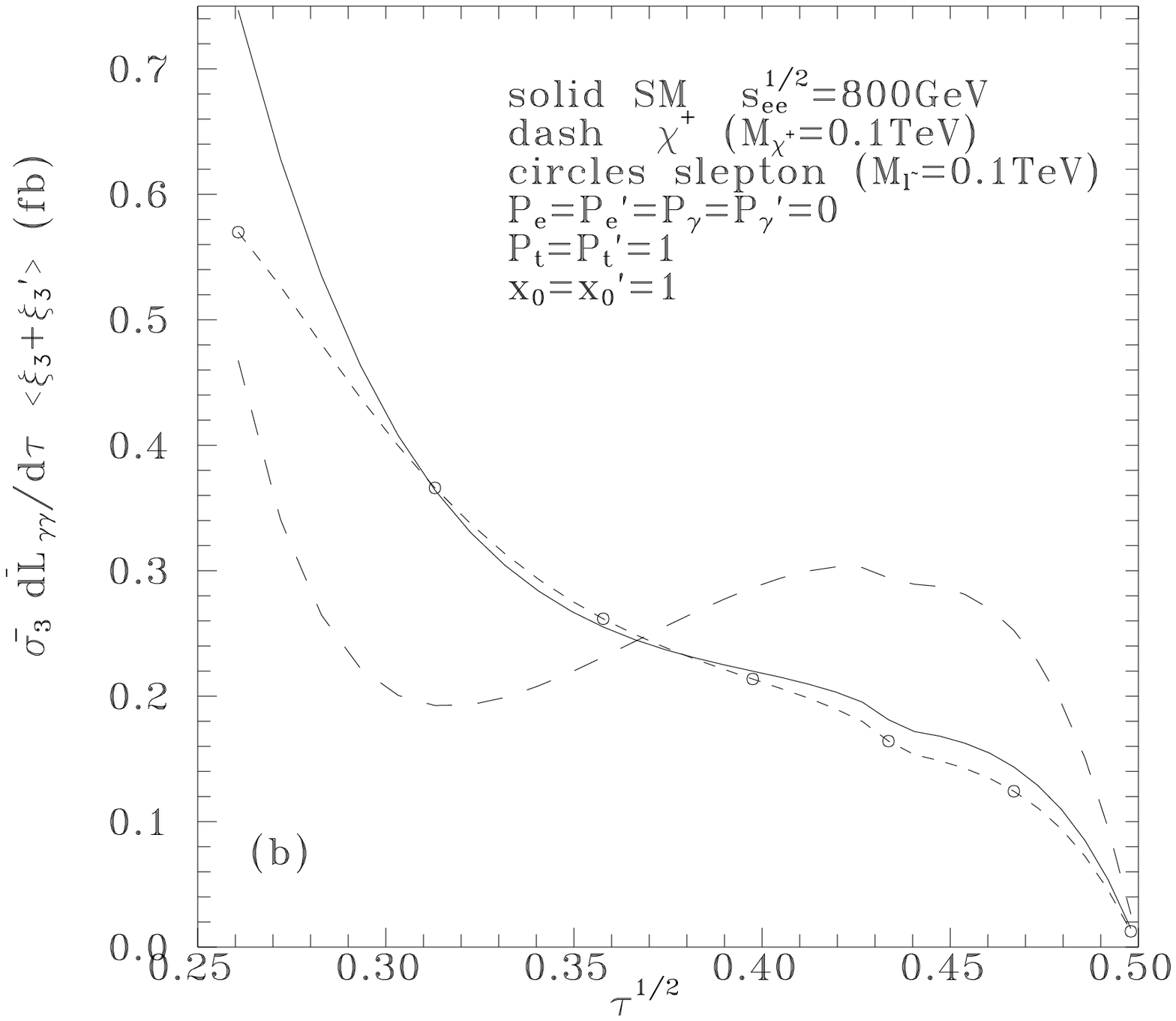,height=7.5cm}
\]
\vspace*{1.5cm}
\[
\epsfig{file=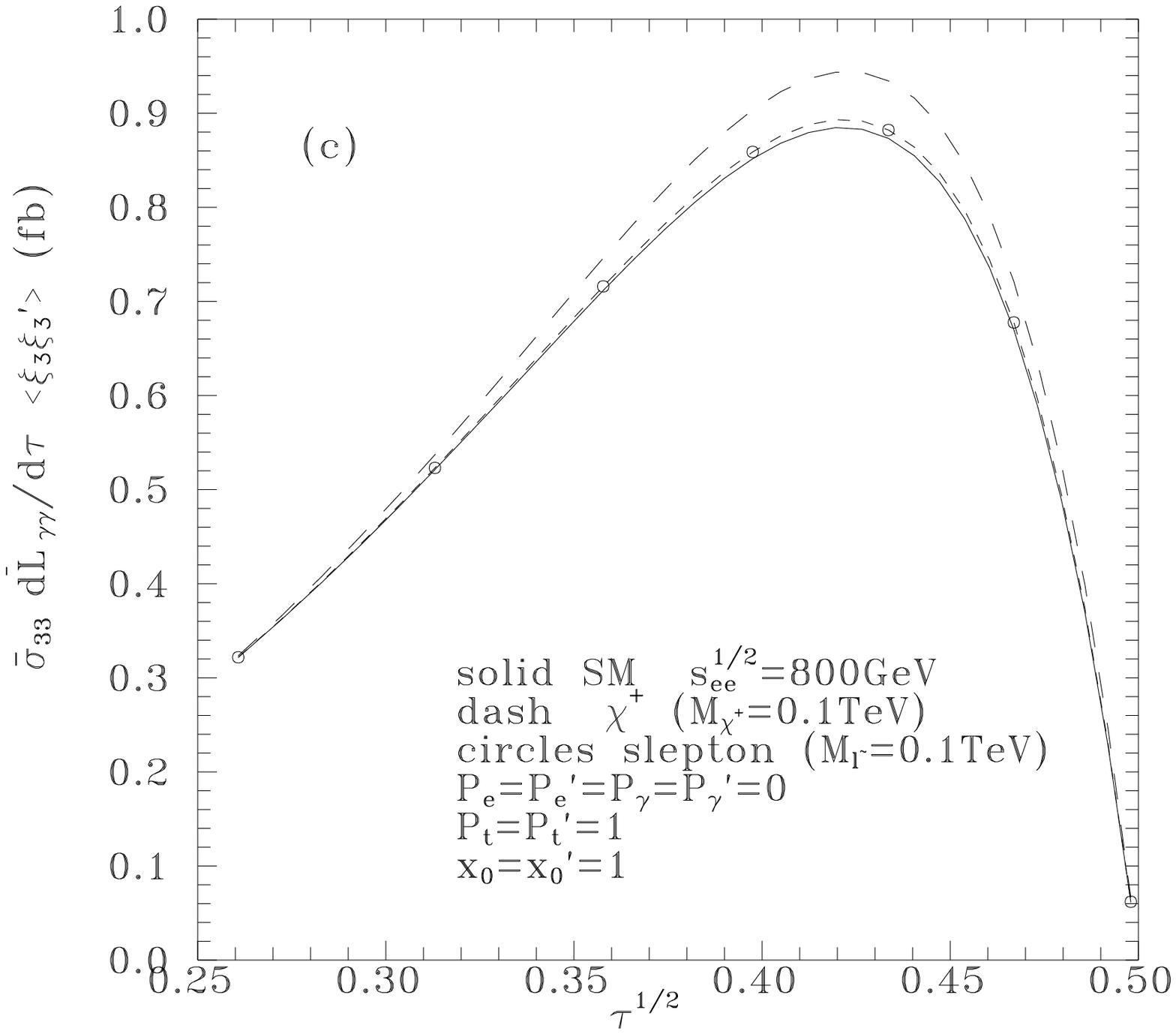,height=7.5cm}\hspace{0.5cm}
\epsfig{file=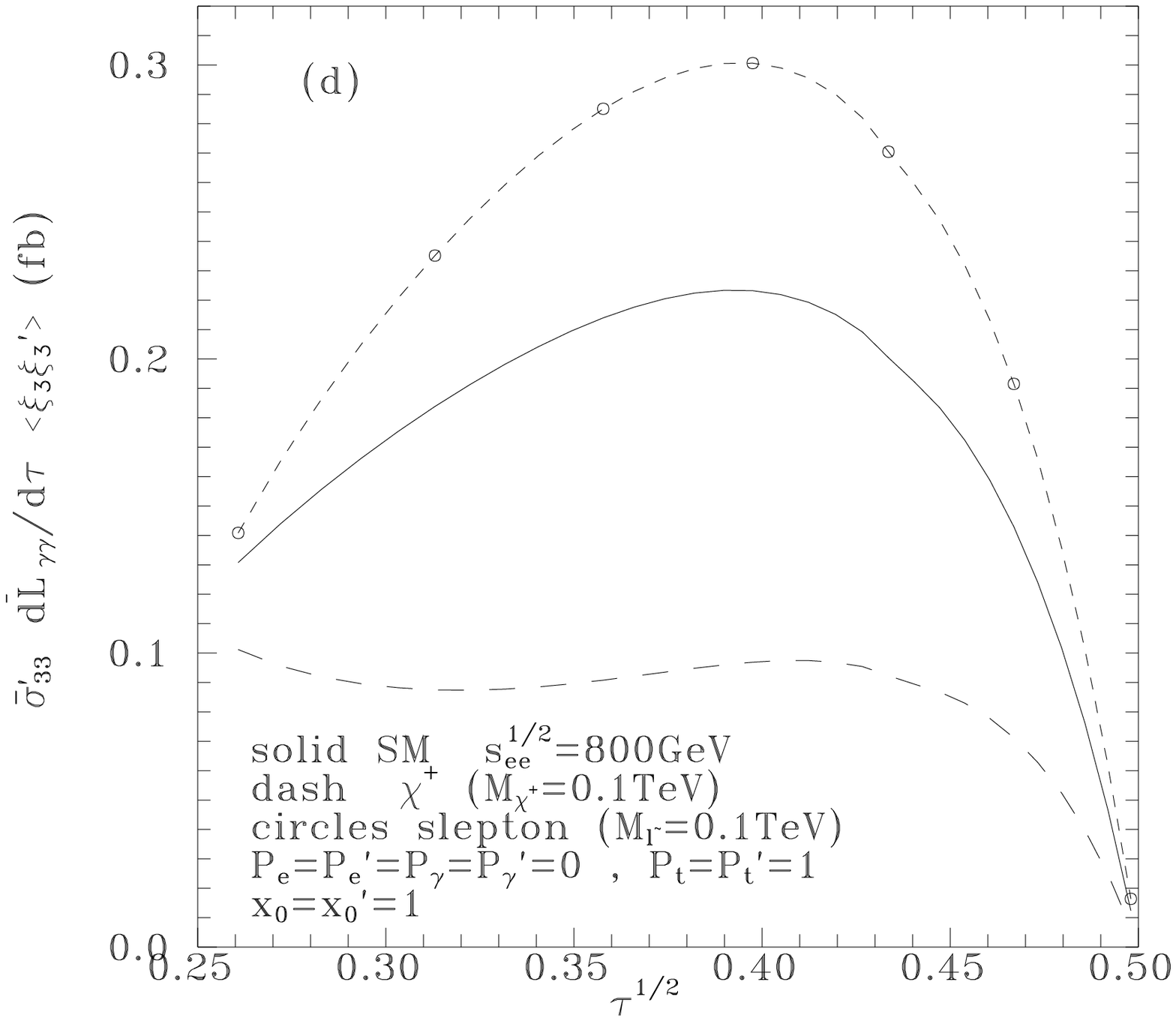,height=7.5cm}
\]
\vspace*{1.cm}
\caption[1]{$\bar \sigma_0$ and  $\bar \sigma_{3}$,
$\bar \sigma_{33}$, $\bar \sigma_{33}^\prime$
cross sections integrated over
$|\cos(\vartheta^*)|<\cos(30^0)$, multiplied by the indicated
photon density matrix elements for the indicated polarizations
 and $x_0$, $x_0^\prime$ parameters.
The SM and SUSY contributions induced by one chargino or
one charged slepton with mass
of 100 GeV, are also indicated.}
\label{SUSY-sig-flux2}
\end{figure}

\clearpage
\newpage

\begin{figure}[p]
\vspace*{-3cm}
\[
\epsfig{file=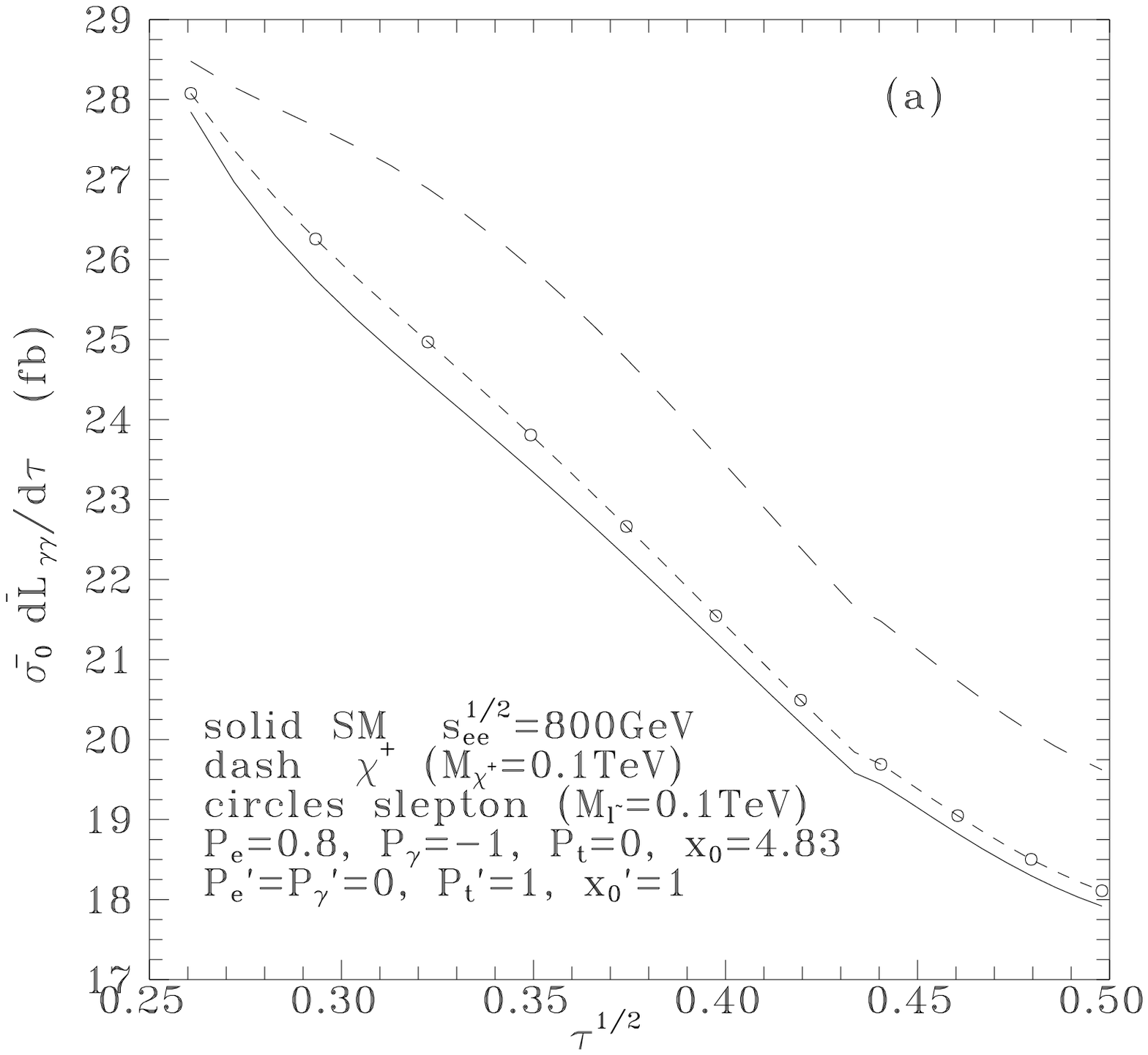,height=7.5cm}\hspace{0.5cm}
\epsfig{file=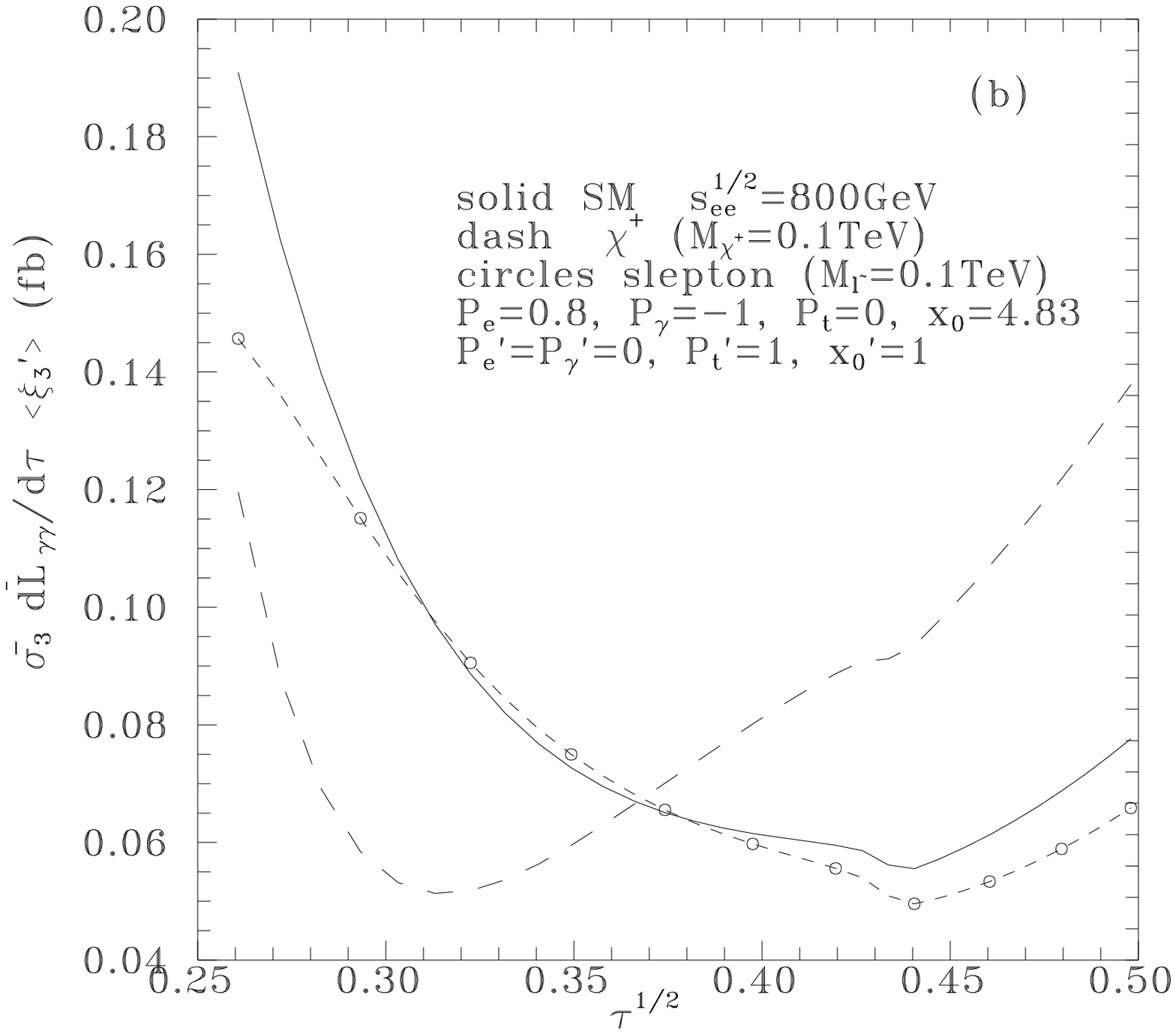,height=7.5cm}
\]
\vspace*{1.5cm}
\[
\epsfig{file=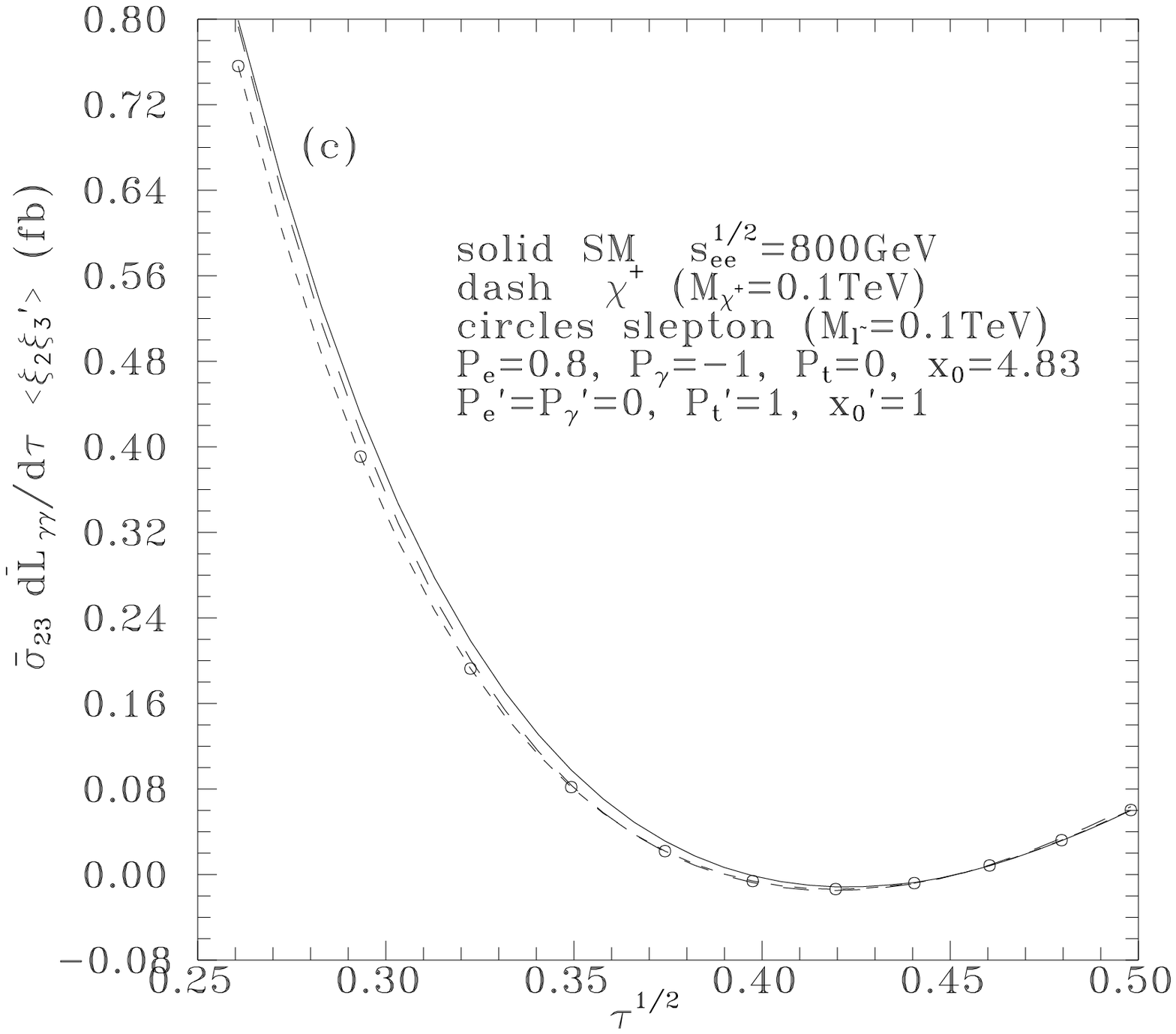,height=7.5cm}
\]
\vspace*{1.cm}
\caption[1]{$\bar \sigma_0$, $\bar \sigma_3$  and
$\bar \sigma_{23}$ cross sections integrated over
$|\cos(\vartheta^*)|<\cos(30^0)$, multiplied by the indicated
photon density matrix elements for the indicated polarizations
 and $x_0$, $x_0^\prime$ parameters.
The SM and SUSY contributions induced by one chargino or
one charged slepton with mass
of 100 GeV, are also indicated.}
\label{SUSY-sig-flux3}
\end{figure}

\end{document}